\documentstyle[graphicx,psfig,epsfig,amsmath,amssymb,enumerate,subfigure]{mn}

\title[Form Filling Functions]
{Cosmological Systematics Beyond Nuisance Parameters : Form Filling Functions}
\author[T. D. Kitching et
  al.]{T. D. Kitching\thanks{tdk@astro.ox.ac.uk}$^{1 {\rm ,}}$$^{2}$, A. Amara$^{3}$, 
F. B. Abdalla$^{4}$, B. Joachimi$^{5}$, A. Refregier$^{6}$ 
\\  
$^{1}$University of Oxford, Department of Physics, Keble Road, Oxford,
OX1 3RH, U.K.\\
$^{2}$SUPA, University of Edinburgh, Institute for Astronomy, Royal
Observatory Edinburgh, Blackford Hill, EH9 3HJ, U.K.\\
$^{3}$Department of Physics, ETH Zurich, Wolfgang-Pauli-Strasse 16, CH-8093 Zurich,
Switzerland\\
$^{4}$Department of Physics \& Astronomy, University College London, Gower Street, London WC1E 6BT, UK.\\
$^{5}$Argelander-Institut fur Astronomie (AIfA), Universitat Bonn, Auf dem Hugel 71, 53121 Bonn, Germany\\
$^{6}$Service d'Astrophysique, CEA Saclay, Batiment 709, 91191 Gif-sur-Yvette Cedex, France\\
}
\newcommand{\be}{\begin{equation}}
\newcommand{\ee}{\end{equation}}
\newcommand{\ba}{\begin{eqnarray}}
\newcommand{\ea}{\end{eqnarray}}
\newcommand{\nn}{\nonumber \\}

\newcommand{\thetab}{\mbox{\boldmath $\theta$}}

\newcommand{\de}{\partial}

\newcommand{\lgl}{\langle}
\newcommand{\rgl}{\rangle}

\newcommand{\bolda}{\mbox{\boldmath $a$}}

\newcommand{\Phib}{\mbox{\boldmath $\Phi$}}

\def\gs{\mathrel{\raise1.16pt\hbox{$>$}\kern-7.0pt %
\lower3.06pt\hbox{{$\scriptstyle \sim$}}}}         %
\def\ls{\mathrel{\raise1.16pt\hbox{$<$}\kern-7.0pt %
\lower3.06pt\hbox{{$\scriptstyle \sim$}}}}         %

\begin{document}


\maketitle


\begin{abstract}
In the absence of any compelling physical model, 
cosmological systematics are often misrepresented as statistical effects and 
the approach of marginalising over extra nuisance systematic parameters 
is used to gauge the effect of the systematic. In this article 
we argue that such an approach is risky at best since the key choice of function can 
have a large effect on the resultant cosmological errors. 

As an alternative we 
present a functional form filling technique in which an unknown, residual, 
systematic is treated as such. Since the underlying function is unknown we  
evaluate the effect of every functional form allowed  
by the information available (either a hard boundary or some data). 
Using a simple toy model we introduce the formalism of functional form 
filling. We show that parameter errors can be 
dramatically affected by the choice of function in the case of 
marginalising over a systematic, but that in contrast the functional form 
filling approach is independent of the choice of basis set.

We then apply the technique to cosmic shear shape measurement systematics and   
show that a shear calibration bias of  $|m(z)|\ls 10^{-3}(1+z)^{0.7}$ 
is required for a future all-sky photometric survey to yield unbiased cosmological 
parameter constraints to percent accuracy.  

A module associated with the work in this paper is available through the 
open source {\tt iCosmo} code available at {\tt http://www.icosmo.org}.
\end{abstract}

\begin{keywords}
Methods: numerical, statistical, data analysis - Cosmology : observation 
\end{keywords}

\section{Introduction}
\label{Introduction}
Cosmology is entering a formative and crucial stage, from a mode in which data sets have 
been relatively small and in which the statistical accuracy required on parameters was 
relatively low, into a regime in which the data sets will be orders of magnitude larger
and the statistical errors required to reveal new physics (for example
 modified gravity -- Heavens et al, 2007, Kunz \& Sapone, 2007;   
massive neutrinos -- Kitching et al., 2008c, Hannestad \& Wong, 2007, 
Cooray, 1999, Abazajian \& Dodelson, 2003; dark energy -- Albrecht et al., 2006, 
Peacock et al., 2006) are smaller than any demanded thus 
far. The ability of future experiments to constrain cosmological parameters 
will not be limited by the statistical power of the probes used 
but, most likely, by systematic effects that will be present in the data, 
and that are inherent to the methods themselves. 

The problem that we will address is how the final level of systematics, 
at the cosmological 
parameter estimation stage, should be treated. 
This problem is of relevance to all cosmological 
probes, some examples include weak lensing and intrinsic alignments (Heavens,
Refregier \& Heymans, 2000; Crittenden et al., 2000; Brown et al., 2002; 
Catelan, et al., 2001; Heymans \& Heavens, 2003; King \& Schneider, 2003; 
Hirata \& Seljak, 2004, Bridle \& King, 2007; Bridle \& Abdalla, 2007); baryon oscillations 
and bias (e.g. Seo \& Eisenstein, 2003), X-ray cluster masses and the 
mass-temperature relation (e.g. Pedersen \& Dahle, 2007) to name a few. 

The general thesis we advocate in this article is that the standard approach to systematics, 
that of assuming some parameterisation and fitting the extra parameters simultaneously to 
cosmological parameters (e.g. 
Kitching et al., 2008a; Bridle \& King, 2007; Huterer et al., 2006; for weak lensing analyses),
both misrepresents a systematic effect as a statistical signal and more
importantly is not robust to the choice of parameterisation. 

As an alternative we will present a method, `form filling functions', 
in which a systematic is treated as such: an 
unknown function which is present in the data. 
By exhaustively exploring the space of functions 
allowed by either data, simulations or theory the effect of a systematic on cosmological 
parameter estimation -- a bias in the maximum likelihood -- can be fully characterised.  
This is a natural extension of the work presented in Amara \& Refregier (2007b).
We will present this using a simple toy model 
to explain and demonstrate the essential aspects of the 
formalism. We will then apply this technique to the problem of shape measurement systematics 
in weak lensing. 

We begin in Section \ref{Alternative Approaches to Systematic Effects} by categorising the 
different approaches to systematics that can be taken, Section \ref{The General Formalism} 
introduces the parameter estimation formalism and how systematics can be included in a 
number of alternative ways. We will then introduce a toy model that
 will then be used to introduce `form filling functions' in Section 
\ref{Functional Form Filling} where we will also compare the standard 
approach to systematics to the one taken here. The application to weak lensing shape measurement systematics is presented in Section \ref{An Application to Cosmic Shear Systematics} and conclusions will be discussed 
in Section \ref{Conclusion}.

\section{Approaches to Systematic Effects}
\label{Alternative Approaches to Systematic Effects}

In this Section we will discuss the problems that may be faced with respect 
to systematics and also address the possible ways that these problems can be addressed.

There are two scenarios in which systematic questions may arise. Either some 
data is available, from 
which information must be extracted, and the effect of systematics on some 
parameters measured 
must be addressed. Or one is planning for a future experiment and the potential impact 
of systematics on some interesting parameters 
must be forecasted. In both cases there may be some extra data available that 
has partially measured the systematic effect, 
or there may be some hard boundary within which 
it is known that the systematic must lie - either from a 
theory or from simulation. 
When forecasting one may want to place a 
constraint on the quality of the extra data needed, 
or the extent of the hard boundary such that future measurements are robust. 
For both data fitting and forecasting there are 
a number of methods that can be employed to 
address the systematics which we review here. 

In the following we will consider a generic method for which the data is an 
observed correlation (covariance) $C^{\rm obs}$ which is a sum of a `signal'    
$C^{\rm signal}(\btheta)$, which depends on a set of statistical (cosmological) 
parameters $\btheta$, and a general additive systematic effect 
$C^{\rm sys}$ (that can, or cannot, depend on the parameters we wish to measure), 
so that the total observed signal is now
\be 
\label{E1}
C^{\rm obs}(\btheta)=C^{\rm signal}(\btheta)+C^{\rm sys}(\btheta)+C^{\rm noise}(\btheta).
\ee
We have also added a benign shot noise term $C^{\rm noise}$ (which again can, or cannot, 
depend on the parameter(s) being measured).  
We do not claim that all systematics can be written this way (but 
most can when the data used is a correlation/covariance of quantities) 
-- a multiplicative bias
is just a special kind of additive term which has the same form as the 
signal but is multiplied by a systematic constant.

We have identified three broad categories of approach that could taken when dealing 
with systematics. 
\\

\noindent{\bf Marginalisation}
\\
Marginalisation of systematics entails using a model, 
a function containing a set of parameters $\bolda$, 
to characterise the systematic effect $C^{\rm sys}\rightarrow C^{\rm sys}(\bolda)$ . 
In this case the cosmological parameters $\btheta$ \emph{and}
the systematic parameters $\bolda$ are measured simultaneously. The extra `nuisance' 
parameter errors are marginalised over to arrive at the final cosmological parameter errors, 
that now take into account the systematic. 

Marginalisation misinterprets the systematic as a statistical signal
(attempts to characterise the systematic by finding best fitting parameters), by reducing the 
estimation and determination of the systematic into a parameter
estimation problem. 
It would be an inappropriate
statistical approach to estimate nuisance parameters that were
known to have a very small degeneracy with cosmological parameters and
then to claim that systematics were negligible.

When marginalising one is immediately faced with the choice of model. 
In the absence of some underlying physical theory one is forced 
to parameterise. The key choice of parameterisation is what makes 
this approach risky (at best); 
both the number of parameters and the prior (if any) on those parameters can dramatically 
affect the level of influence that the systematic may have on 
cosmological parameter estimation.
One can choose either simple models, whose small degree of freedom may have a 
minimal impact on the cosmological parameters, but whose behaviour may mask the true 
systematic signal. Or, very flexible models; but one is always limited
by the number of degrees of freedom that can be estimated from the
data, and using for example $\gg 100$ nuisance parameters to find the
systematic error on $\sim 10$ cosmological parameters seems assymetric.

In some circumstances there are physical models that can be called upon to model a systematic 
accurately, in this case marginalisation becomes an attractive option. 
One could also use more sophisticated techniques, such as Bayesian evidence, 
to determine which 
parameterisation is warranted given the data available. But even in such a scenario the 
question of 
whether an \emph{even more} apt model is available, or not, would always remain and 
even in this case a \emph{residual} systematic will 
remain (at least due to noise) which may contain a still unknown
effect and must be treated in the correct way.
\\

\noindent {\bf Bias Formalism}
\\
The systematic is not marginalised over in cosmological parameter estimation, 
but is left in as a systematic term $C^{\rm sys}\not= C^{\rm sys}(\btheta)$.  
By `leaving a systematic in' and not marginalising over parameters 
the systematic is correctly identified as a systematic effect, albeit that 
the magnitude of the effect must be correctly quantified.  
If a systematic is `left in' then the cosmological parameter errors themselves 
are unaffected (in the case that the observation is not dominated by the systematic). 
The maximum likelihood value of the cosmological parameters however 
will always be biased by an amount which 
depends upon the true, underlying, systematic signal. There have been studies of the 
biases that can be caused when a systematic is treated as such (e.g. Huterer \& Takada, 2005; 
Amara \& Refregier, 2007b; Kitching et al., 2008a) 
although all studies have assumed some 
functional form for the underlying systematic.

The task is then to investigate 
all possible functional forms for the systematic, 
that are allowed by either theory or data so quantifying the extent of the possible biases. 
In this case flexibility is paramount since every 
possible allowed function must be tested. This is the approach advocated in this article.

By treating the systematic in this way, as a true systematic effect as 
opposed to a statistical effect (as in marginalisation), we can move away from the dilemma of 
choosing a particular parameterisation.
\\

\noindent{\bf Nulling}
\\
The general approach to nulling is that the statistical signal used (the way in which
the data is used to extract cosmological information) can be modified in 
such a way that the systematic signal is cancelled out i.e. 
$C^{\rm obs} = C^{\rm signal} + C^{\rm sys} + C^{\rm noise}\rightarrow C^{\rm obs}_{\rm new} 
= C^{\rm signal}_{\rm new}  + C^{\rm noise}_{\rm new}$ and $C^{\rm sys}_{\rm new}=0$. 
Cosmological parameter estimates 
can then be made using this new statistic which by construction has minimised, 
or completely removed, the systematic effect.  

The nulling approach is a potentially powerful tool, for 
example as shown by Joachimi \& Schneider (2008) 
this could be used in 
the removal of weak lensing intrinsic alignment contaminant. 
However when nulling, the cosmological 
signal is changed in such a way that parameter constraints can be severely degraded.
We will not address the 
nulling approach further in this article but we note that the possibility 
of `optimal' weighting (partial nulling)
should exist, in a mean-square error sense. Nulling aims to set the bias due to a 
systematic to zero, which 
may be to strict since our true requirement 
is simply that the biases are sub-dominant to the statistical errors.
\\
\\


In the next Section we will review 
the marginalisation procedure and formalise the biasing effect of 
leaving a systematic in the signal. What we 
endorse within the context of the bias formalism  
is using the theory and data itself to investigate the full range of 
allowed functions, thus fully characterising the effect 
that a systematic may have.

\section{The General Formalism}
\label{The General Formalism}
A common approach in cosmology is to measure the signal of some quantity, 
and match this to 
theory in order to constrain cosmological parameters.
However, as we will show,  
the effect of systematics on such cosmological probes is usually dealt with in a way 
which can potentially mask their true impact. 

\begin{figure}
\resizebox{84mm}{!}{
\includegraphics{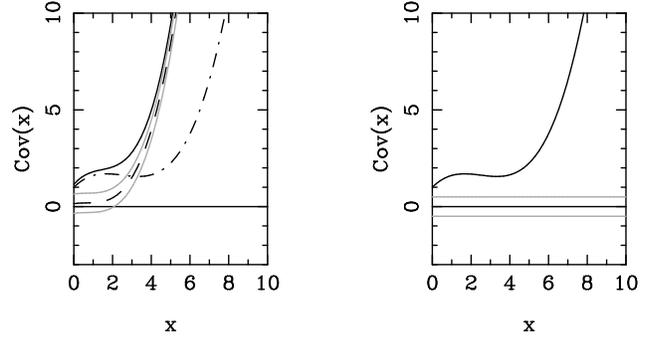}}
\caption{An example of the basic premise concerning the parameter estimation methodology. 
The left panel shows a total observed correlation (solid black line), that is a sum 
of the signal (dot dashed line) and systematic plus noise (dashed line). 
The systematic is known to lie within some tolerance envelope (within the gray solid lines). 
The right panel shows the observation minus the mean of the systematic plus noise, 
leaving an estimator of the signal (solid line) and a systematic tolerance about zero.}
\label{basic}
\end{figure}

Fig. \ref{basic} shows the basic situation which we will address. 
The left panel shows the 
observable $C^{\rm obs}$ which is a sum of the signal $C^{\rm signal}$ 
and some 
systematic plus noise $C^{\rm sys}+C^{\rm noise}$, equation (\ref{E1}). 
Furthermore there is some tolerance 
envelope around the systematic (gray solid lines) which represents the state of knowledge
regarding the systematic. One can then subtract the mean $C^{\rm sys}$ and $C^{\rm noise}$ 
from the observable which results in an estimator of the signal $\widehat{C^{\rm signal}}$ 
\be
\label{E2}
C^{\rm obs}-\langle C^{\rm sys}\rangle-\langle C^{\rm noise}\rangle=
\widehat{C^{\rm signal}}+\widetilde{C^{\rm sys}}
\ee
plus some residual systematic $\widetilde{C^{\rm sys}}$ which is
centered around zero. In general throughout we always consider the
case that there is some extra data that places constraints on the
systematic $\widetilde{C^{\rm sys}}$.
This is shown in the right panel of Fig. \ref{basic}, the systematic tolerance envelope 
now lies about the $C(x)=0$ line. 

The measurement of this signal to estimate the values of some 
parameters $\btheta$ within a theory $C_{\rm theory}^{\rm signal}(\btheta)$ can be done 
in the usual way  
\be
\label{E3}
\chi^2(\btheta)=\sum_x \sigma^{-2}_C[\widehat{C^{\rm signal}}-C_{\rm theory}^{\rm signal}(\btheta)]^2
\ee
where  $\sigma_C(x)$ is the error on the signal. 
Note that we will remove $(x)$ (e.g $\sigma_C(x)\rightarrow\sigma_C$) 
from all equations for clarity. 
A best estimator of the parameters from the observation $\hat\btheta$ 
is defined such that $d\chi^2/d\btheta=0$. 
However this statistic has not taken into account the 
residual systematic effect in any way. 

To make predictive statements regarding parameter estimation it is convenient to work with 
the Fisher matrix formalism. The Fisher matrix allows for 
the prediction of parameter errors given a specific experimental design and 
method for extracting parameters. In the case of Gaussian-distributed
data where we assume that the error on the signal is not a function of parameter values 
$\sigma_C\not=\sigma_C(\btheta)$ we can take the covariance 
of the estimated values of the parameters 
(Tegmark, Taylor \& Heavens, 1997; Jungman et al., 1996; Fisher, 1935)
\be 
{\rm cov}[\hat\btheta_i,\hat\btheta_j]=\langle (\hat\btheta_i-\langle\hat\btheta_i\rangle)
(\hat\btheta_j-\langle\hat\btheta_j\rangle)\rangle=F^{-1}_{ij}
\ee
where the Fisher matrix is defined by 
(Tegmark, Taylor \& Heavens, 1997; Jungman et al., 1996; Fisher, 1935)
\be
\label{fishe}
F_{ij}=\sum_x \left[\sigma^{-2}_C\frac{\partial C}{\partial \btheta_i}
\frac{\partial C}{\partial \btheta_j}\right].
\ee
The marginal errors on 
the parameters are given by $\Delta\theta_i=\sqrt{(F^{-1})_{ii}}$, 
this is the minimum marginal error that one can expect for the experimental 
design considered (due to the Cramer-Rao inequality; Tegmark, Taylor \& Heavens, 1997). 
\begin{table*}
\begin{center}
\begin{tabular}{|l|c|c|c|}
\hline
Approach&Broadens Likelihood&Bias Likelihood&Problems\\
\hline
Marginalise&$\surd$&$\times$&
Choice of parameterisation\\
Bias Formalism&$\times^{\dagger}$&$\surd$&
Need to assess all allowed functions\\
\hline
\end{tabular}
\caption{A summary of the different approaches to systematic effects, 
  showing the effect on the likelihood surface and the primary problem that 
  each method encounters. 
  $^{\dagger}$ Only in the case that the systematic is sub-dominant to the signal.}
\label{categorisation}
\end{center}
\end{table*}

\subsection{Model Fitting and Marginalisation} 
\label{Model Fitting and Marginalisation}
The marginalisation approach fits a model to the residual systematic and treats 
the systematic as an 
extra statistical effect. The model chosen for the systematic 
$C_{\rm theory}^{\rm sys}(\bolda)$ 
depends on a suite of new parameters 
$\bolda$ and on the original parameter set $\btheta$ where the total parameter set is given by
$\Phib=(\btheta,\bolda)$.
The extra parameters are then assumed to be part of the signal of a method. 
To estimate the values of the parameters $\btheta$ the $\chi^2$ statistic 
of equation (\ref{E3}) is modified to 
\ba
\label{E4}
&&\chi^2_{\rm total}(\btheta,\bolda)=\nn
&&\sum_x \sigma^{-2}_{C^{\rm signal}}[\widehat{C^{\rm signal}}-
C_{\rm theory}^{\rm signal}(\btheta)-C_{\rm theory}^{\rm sys}(\bolda)]^2+\nn
&&\sum_x \sigma^{-2}_{C^{\rm sys}}[\widetilde{C^{\rm sys}}-
C_{\rm theory}^{\rm sys}(\bolda)]^2
\ea
where the total $\chi^2$ is minimised to find the best estimator of the parameters. 
The likelihood functions for the cosmological parameters are found by 
marginalising the combined likelihood $p(\btheta,\bolda)$ over the new parameters
\be 
p(\btheta)=\int d\bolda p(\btheta,\bolda).
\ee

The new Fisher matrix for the total parameter set
$\Phib$ becomes a
combination of the cosmological Fisher matrix $F^{\btheta\btheta}$, 
the derivatives of the likelihood
with respect to the cosmological parameters and the systematic parameters 
$F^{\btheta\bolda}$ and the
systematic parameters with themselves $F^{\bolda\bolda}$
\be
\label{fishf}
F^{\Phib\Phib}=
\left( \begin{array}{cc}
 F^{\btheta\btheta} & F^{\btheta\bolda}  \\
F^{\bolda\btheta}   & F^{\bolda\bolda} \\
  \end{array}\right).
\ee
Where the individual terms are given by
\ba 
\label{E5}
F^{\btheta\btheta}_{ij}&=&\sum_x \left[\sigma^{-2}_{C^{\rm signal}}\frac{\partial C_{\rm theory}^{\rm signal}}{\partial \btheta_i}
\frac{\partial C_{\rm theory}^{\rm signal}}{\partial \btheta_j}\right]\nn
F^{\btheta\bolda}_{ij}&=&\sum_x \left[\sigma^{-2}_{C^{\rm signal}}\frac{\partial C_{\rm theory}^{\rm signal}}{\partial \btheta_i}
\frac{\partial C_{\rm theory}^{\rm signal}}{\partial \bolda_j}\right]\nn
F^{\bolda\bolda}_{ij}&=&\sum_x \left[\sigma^{-2}_{C^{\rm signal}}\frac{\partial C_{\rm theory}^{\rm sys}}{\partial \bolda_i}
\frac{\partial C_{\rm theory}^{\rm sys}}{\partial \bolda_j}\right]\nn
&+&\sum_x \left[\sigma^{-2}_{C^{\rm sys}}\frac{\partial C_{\rm theory}^{\rm sys}}{\partial \bolda_i}
\frac{\partial C_{\rm theory}^{\rm sys}}{\partial \bolda_j}\right]
\ea
here we have assumed that the errors are uncorrelated and do not depend on the parameters.  

The predicted 
cosmological parameter errors now including the effect of the systematic are given by 
$\Delta\theta_i=\sqrt{[(F^{\Phib\Phib})^{-1}]_{ii}}$ (see Appendix A for a more 
detailed expression). The cosmological parameter errors are 
increased due to the degeneracy between the cosmological and 
systematic parameters (included by the $F^{\btheta\bolda}$ terms). 
The tolerance envelope around the residual systematic (Fig. \ref{basic}) acts as  
a prior on the systematic parameters in the chosen model. 

\subsection{The Bias Formalism}
\label{The Bias Formalism}
The bias formalism treats a systematic as such, 
by not statistically marginalising over any extra 
parameters within a model. Instead the systematic simply adds an extra systematic function 
to the signal. By doing this a bias is introduced in the maximum likelihood
value of the parameters with respect to the true underlying values
\be 
b[\btheta_i]=\langle\hat\btheta_i\rangle-\langle\hat\btheta^{\rm true}_i\rangle.
\ee
When marginalising the choice lies in the suite of parameters, and the function, chosen. Here 
there is a similar choice, one must assume that the systematic has some functional form 
$C_{\rm function}^{\rm sys}$.
To estimate the values of the cosmological parameters $\btheta$, 
the $\chi^2$ statistic of equation (\ref{E3}) is modified to include the assumed systematic  
\be
\label{E6}
\chi^2(\btheta)=\sum_x \sigma^{-2}_C [\widehat{C^{\rm signal}}+C_{\rm function}^{\rm sys}-
C_{\rm theory}^{\rm signal}(\btheta)]^2.
\ee
The estimate of the parameter values will now be biased 
but the marginal error on the parameters
will remain the same (the caveat here that this is only the case when the systematic is smaller than the signal).

It can be shown (Taylor et al., 2007; Amara \& Refregier, 2007b; Kim et al., 2004) that, 
with the assumption of Gaussian likelihoods, 
the predicted bias in a parameter due to an uncorrected systematic is given by 
\be 
\label{biase}
b[\btheta_i]=(F^{-1})_{ij}B_j
\ee
where  
\ba 
F_{ij}=F^{\btheta\btheta}_{ij}&=&\sum_x \left[\sigma^{-2}_{C}\frac{\partial C_{\rm theory}^{\rm signal}}{\partial \btheta_i}
\frac{\partial C_{\rm theory}^{\rm signal}}{\partial \btheta_j}\right]\nn
B_j&=&\sum_x\sigma^{-2}_C C_{\rm function}^{\rm sys}\frac{\partial C_{\rm theory}^{signal}}{\partial\btheta_j}.
\ea

To recap Sections \ref{Alternative Approaches to Systematic Effects} and 
\ref{The General Formalism}, Table \ref{categorisation} 
summarises the effect on the likelihood surface and the primary problem 
encountered by each systematic approach. 

In cases where the systematic affects the error on the signal  
$\sigma_C=\sigma_C(C^{\rm sys})$ (which is almost always) 
then leaving a systematic in the signal can cause a bias and increase the marginal errors. 
However the increase in marginal errors is negligible 
for systematics that have an amplitude which is much less than the signal, 
and cause biases 
that are $\ls 10\sigma$ (Amara \& Refregier, 2007). 
Equation (\ref{biase}) is also an approximation 
for the case of small biases, if the bias is large relative to the marginal error 
then the curvature of the likelihood surface will have varied substantially from the 
Gaussian approximation. In such cases one could go to a higher order in the 
Taylor expansion used to derive equation (\ref{biase}), or calculate the full likelihood.

\subsection{A Simple Example} 
\label{A Simple Toy Model}
To review the general formalism described thus far,  
and for use in subsequent Sections, we will here introduce 
a simple model. 
The toy model we will consider is shown in the right hand panel of Fig. \ref{basic}. Referring 
to equation (\ref{E1}) the signal is given by a simple polynomial expansion
\be
\label{ex1} 
C^{\rm signal}_{\rm example}(x)=a_0+a_1x+(-0.45)x^2+(0.05)x^3
\ee
where the statistical parameters we are concerned with (with fiducial, true, values) 
are the parameters $a_0=1.0$ and $a_1=1.0$. We assume that 
the observed signal is measured with an error of $\sigma_C(x)=0.25$. We include 
the $x^2$ and $x^3$ terms so that the problem is slighly more 
realistic, in that there is an extra behaviour in the signal that we do not wish 
to constrain but may effect the determination of the parameters of interest.

\begin{figure}
\resizebox{84mm}{!}{
\includegraphics{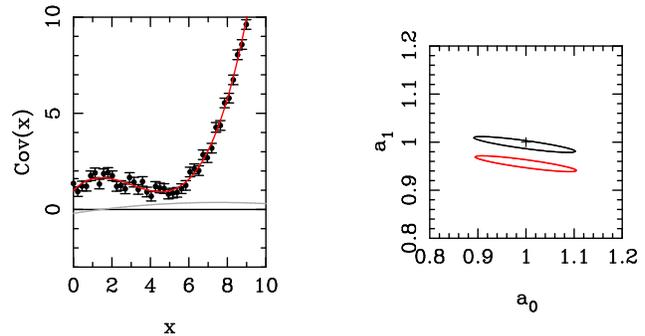}}
\caption{The left panel shows the toy model signal, the fiducial model is shown in red 
(dark gray) and we have added some simple Gaussian distributed data points about this 
central model. We also show an example systematic in (light) gray defined 
in equation (\ref{ex2}). The right 
panel shows the two-parameter $1$-$\sigma$ error contours without a systematic (black) and 
including the example systematic (red/dark gray). The fiducial model is marked by $+$.}
\label{bias_example}
\end{figure}

Fig. \ref{bias_example} shows the simple model signal with some Gaussian distributed 
data points. We then calculate the two-parameter marginal errors 
using equation (\ref{E3}) where
we use the model as in equation (\ref{ex1}). Table \ref{extab} shows the measured marginal 
errors from this simple mock data, and compares these with the expected marginal errors 
calculated using the Fisher matrix (equation \ref{fishe}).

We then introduce a simple systematic into the model by assuming a simple function 
(note this does not have a mean of zero, but could fit into some boundary centered on $C(x)=0$)
\be
\label{ex2} 
C^{\rm sys}_{\rm example}(x)=-0.2+0.15x-0.01x^2.
\ee
By recalculating the likelihood and including this systematic function, 
using equation (\ref{E6}), we find that the most likely value of $a_0$ and $a_1$ 
is biased and yet the marginal errors remain the same. 
We compare this bias to the prediction made using equation (\ref{biase}) in Table \ref{extab}.

\begin{table}
\begin{center}
\begin{tabular}{|l|c|c|}
\hline
\hline
{\bf No Systematics}\\
Parameter&Measured Error&Expected Error\\
\hline
$a_0$&$0.100$&$0.099$\\
$a_1$&$0.011$&$0.011$\\
\hline
\hline
{\bf Bias Method}\\
\hline
Parameter&Measured Bias&Expected Bias\\
$a_0$&$0.003$&$0.002$\\
$a_1$&$0.040$&$0.040$\\
\hline
\hline
{\bf Marginalising Method}\\
\hline
Parameter&Measured Error&Expected Error\\
$a_0$&$0.120$&$0.119$\\
$a_1$&$0.012$&$0.013$\\
\hline
\end{tabular}
\caption{This table compares the Fisher matrix predictions with values found using some simple 
mock data described in Section \ref{A Simple Toy Model}. This is not a comparison of the 
systematic methods themselves, which will be done in Section \ref{Marginalisation_Bias}. 
The upper table shows the marginal errors found using 
the data shown in Fig. \ref{bias_example} for the parameters $a_0$ and $a_1$ 
defined in equation (\ref{ex1}). These errors are compared to what is expected from the 
Fisher matrix, equation (\ref{fishe}). The middle table shows the measured bias 
when a simple systematic is added (equation \ref{ex2}) and compares this to the expected 
bias calculated using equation (\ref{biase}). 
The lower table shows the increased marginalised errors on $a_0$ and $a_1$ when a simple 
parameterised systematic model is marginalised over.}
\label{extab}
\end{center}
\end{table}

\begin{figure}
\resizebox{84mm}{!}{
\includegraphics[clip=true]{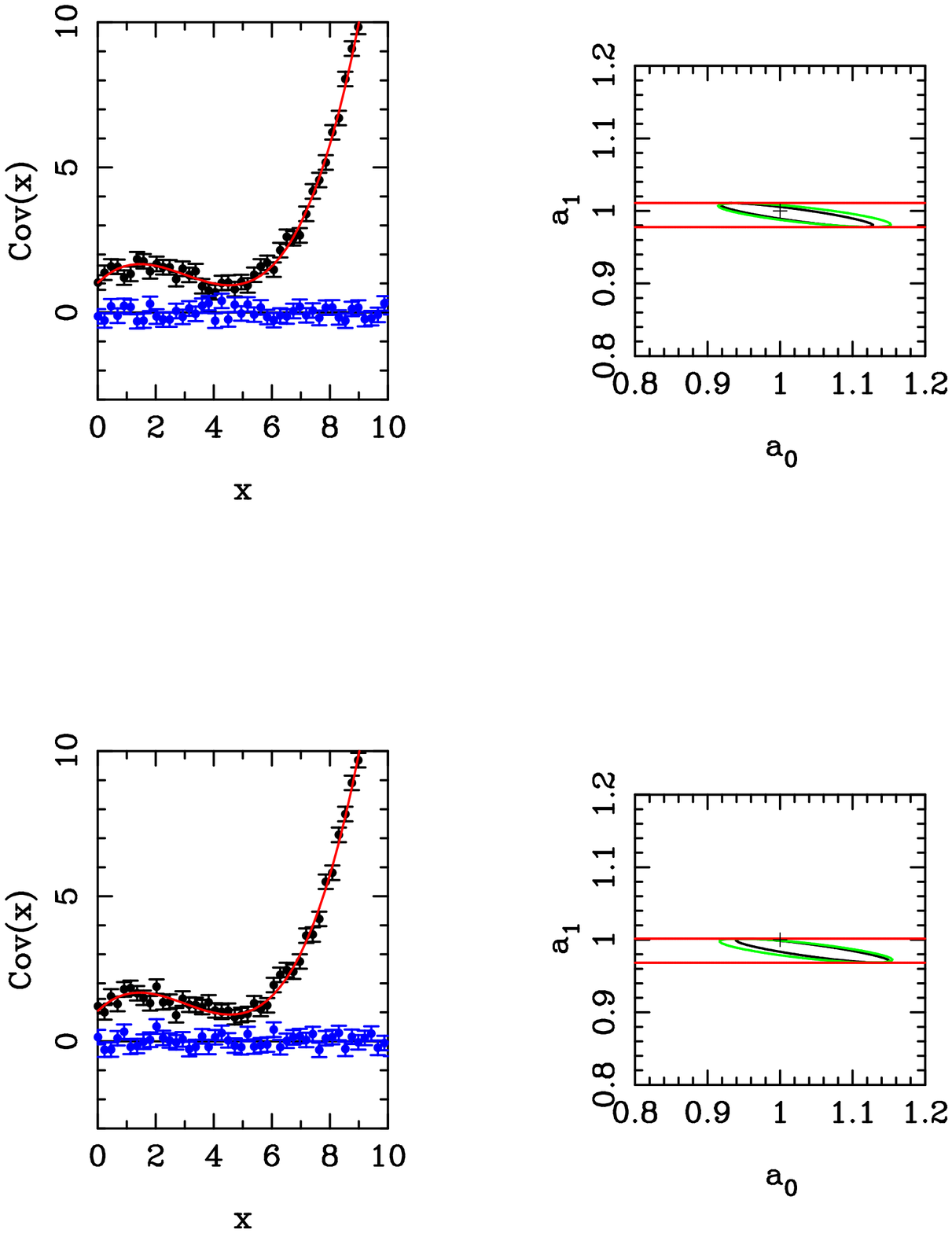}}
\caption{The left panel shows the model signal, the fiducial model is shown in red 
(dark gray) and we have added some Gaussian distributed data points about this 
central model. We also show an example data-driven systematic blue (light gray) points about 
$C(x)=0$ with a variance of $\sigma_{C^{\rm sys}}=0.5$. The right 
panel shows the two-parameter $1$-$\sigma$ error contours with (green/light gray) and 
without (black) marginalising over 
the systematic model. The red lines show the marginal error with no
extra systematic data, a complete degeneracy between $a_0$ ans $s_0$. 
The fiducial model is marked by $+$.}
\label{marg_example}
\end{figure}

To test the method of marginalising over data we now reset the
systematic (throw away equation, \ref{ex2}) and introduce a 
simple systematic which is measured
using some mock data. Fig. \ref{marg_example} shows the model signal with some extra systematic data with a mean of zero and a scatter of $\sigma_{C^{\rm sys}}=0.5$. We then 
introduce a simple systematic model parameterised by a new parameter $s_0$ 
\be 
C^{\rm sys}_{\rm theory\, example}(x)=s_0
\ee
and fit the model (equation \ref{ex1}) and the systematic to the data simultaneously, as 
described in equation (\ref{E4}). The total likelihood $p(a_0,a_1,s_0)$ is then 
marginalised over $s_0$. Fig. \ref{marg_example} shows that when this 
extra systematic is marginalised over the constraint on $a_0$ is affected the most since 
$s_0$ has the same functional form as this parameter, and so there exists a
large degeneracy between $a_0$ and $s_0$. The exact degeneracy is broken by the extra data 
available for the systematic. In Table \ref{extab} we show the increase in 
the marginal error on the parameters $a_0$ and 
$a_1$ when we marginalise over the extra systematic parameter $s_0$.

We have now introduced the basic formalism and shown that this can be applied to a simple 
example that yields results which are in good agreement with the Fisher matrix predictions. 

\section{Form Filling Functions} 
\label{Functional Form Filling}

For the remainder of this article we will present an alternative to marginalisation by 
advocating the bias formalism for dealing with 
systematics, outlined in Section \ref{The Bias Formalism}. 
The issue with which one is now faced is what function to choose for the residual 
systematic. 
To investigate the full extent of possible biases, allowed by the tolerance on the systematic,
all allowed functions must be addressed in some way.

\begin{figure}
\resizebox{40mm}{!}{
\includegraphics[clip=true]{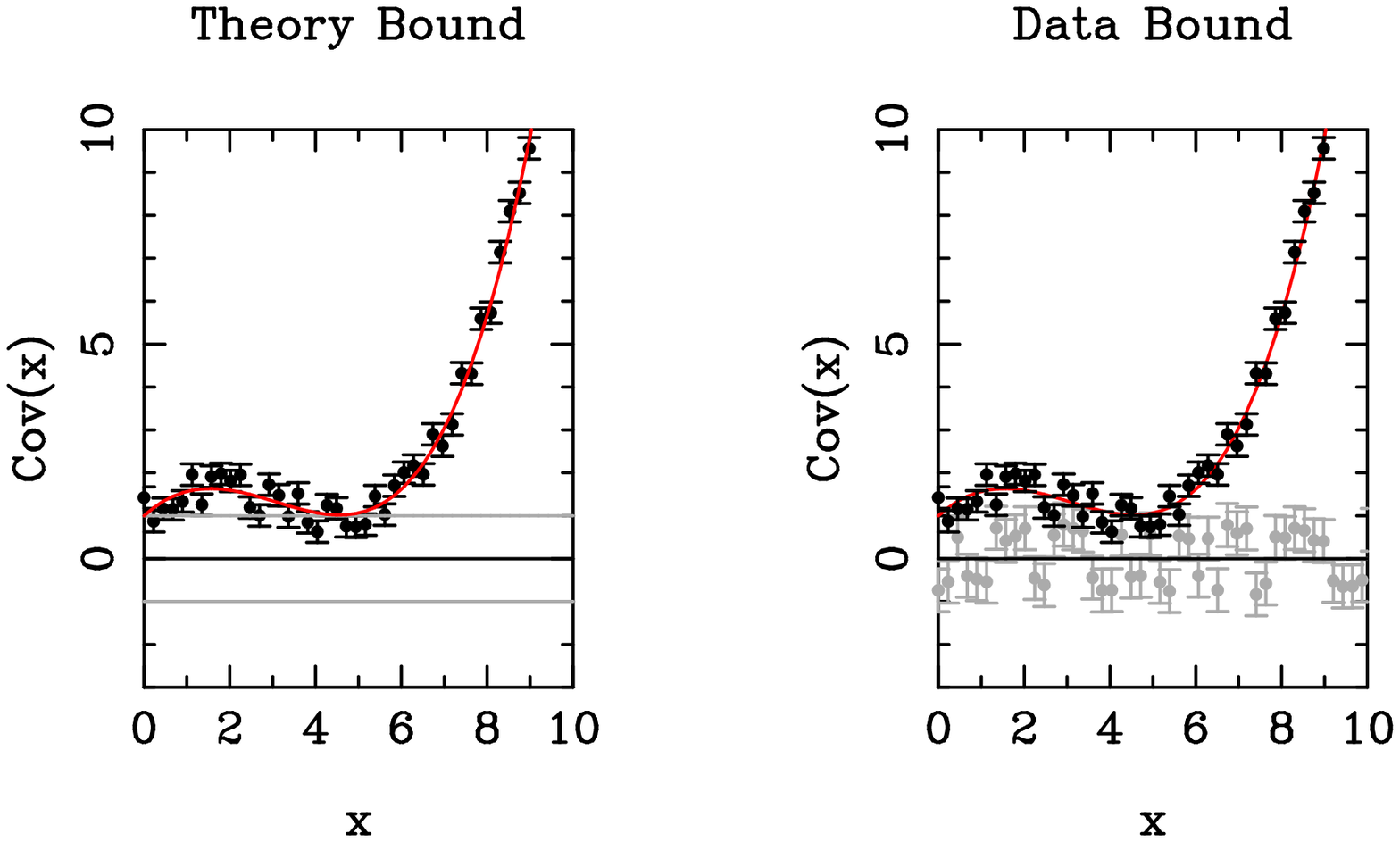}}
\resizebox{41mm}{!}{
\includegraphics[clip=true]{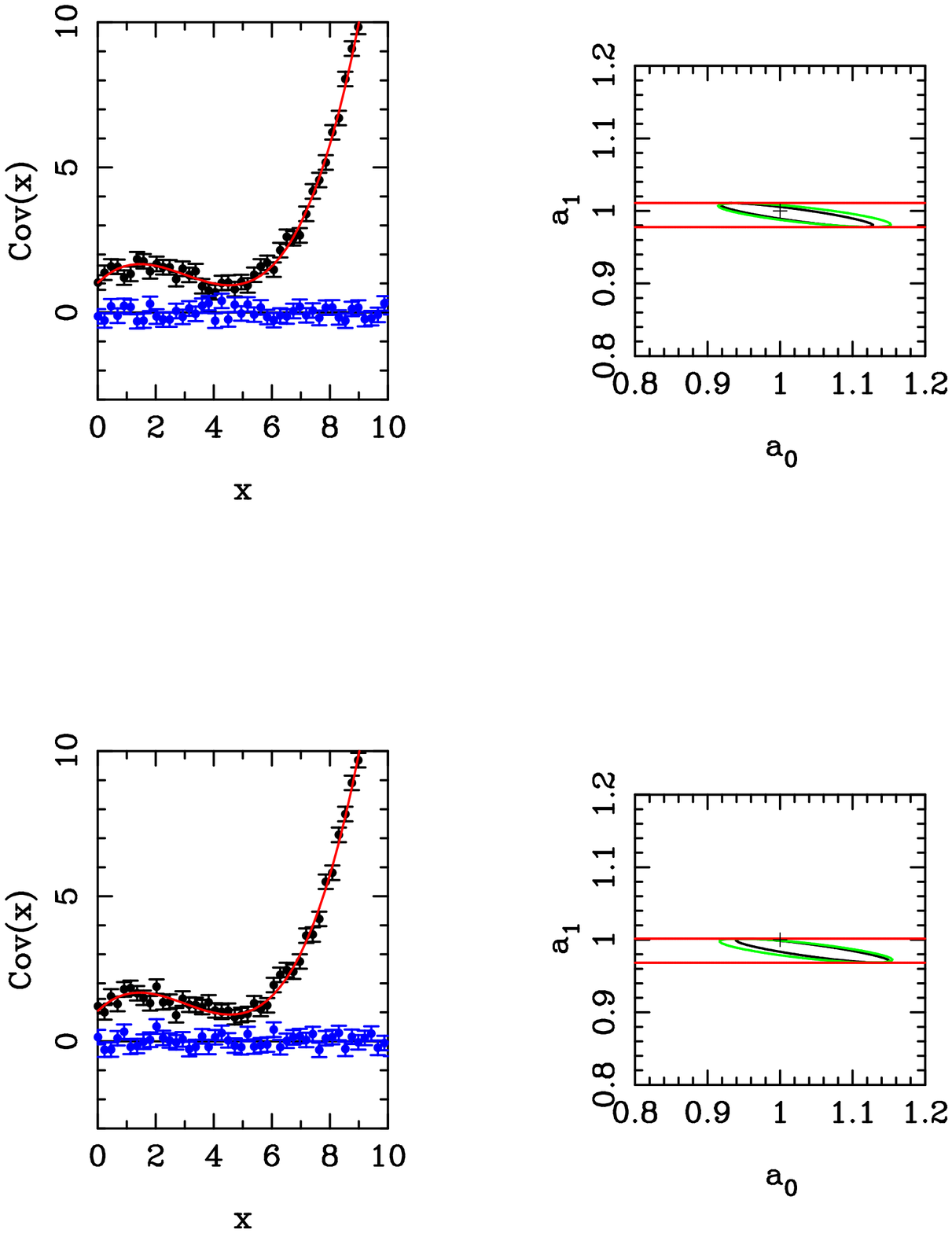}}
\caption{Representing the two possible systematic constraints, 
either from theory or from data. 
A hard boundary (left panel solid gray lines) may be defined 
within which the systematic must lie, or some data may 
provide a measurement of the level of systematic (right panel
blue/light gray error bars). The black data 
points are mock data with a Gaussian distribution using the model described in Section 
\ref{A Simple Toy Model}, the red (dark gray) line is the fiducial model.}
\label{theory_data}
\end{figure}

In Fig. \ref{theory_data} we use an extension of the simple model, 
outlined in Section \ref{A Simple Toy Model}, 
to introduce the concept of two different forms of systematic 
tolerance. The tolerance 
envelope could have a \emph{hard boundary} (e.g. defined by a theory which states that ``the 
systematic \emph{must} lie within this boundary''). 
Or the systematic could be defined by some extra data 
that has partially measured the magnitude of the systematic. We will refer to these two 
scenarios as the ``hard bound'' and ``data bound'' respectively.

To fully assess the level of 
bias every functional form allowed by the 
systematic tolerance envelope needs to be tested. For the hard boundary we want to find 
every function that can be drawn within the hard boundary. For the data bound every function 
can be weighted with respect to the data itself. Here we 
introduce the concept of `form filling functions' which are a set of functions that should 
exhaustively fill the space of possible functions allowed by some tolerance envelope. 

Consider the hard bound in Fig. \ref{theory_data}, in the bias approach 
we want to find every function that will 
fit within this tolerance envelope. 
To do this we consider functions in the most general form as expansions 
in some arbitrary basis set
\be 
\label{E7}
f(x; \{a_n\}, \{b_n\})=\sum^N_{n=1} a_n \psi_n(x) + b_n \phi_n(x)
\ee
where $a_n$, $b_n \in \Re$ and $\psi_n(x)$ and $\phi_n(x)$ form some arbitrary 
basis functions. 
To choose the basis set we impose the following conditions
\begin{itemize} 
\item 
The basis set must be \emph{complete} in the range of $x$ we are considering i.e. all functions $f(x)$ must be expressible as an expansion in the basis.
\item 
The basis functions must be orthogonal.
\item 
The functions must be \emph{boundable} i.e. the basis set must be able to be manipulated such that every function described within some bounded region (tolerance envelope) can be 
drawn.
\end{itemize}
The first condition is necessary. The second condition is make some calculations more 
straightforward though is not strictly necessary. The third condition 
is merely desirable -- one can imagine having a basis set in which 
non-bounded functions are allowed, but when using 
such expansions one would have remove these stray functions. 
As an aside we note that over-completeness is not a problem as long as the basis set is in 
fact complete (we do not care if we sample a function multiple times, as long 
as we sample it at least once)\footnote{Orthogonal basis sets are
  never over-complete so the second condition means we shouldn't be in
  this situation for our functional form filling algorithm.}. 
\begin{figure*}
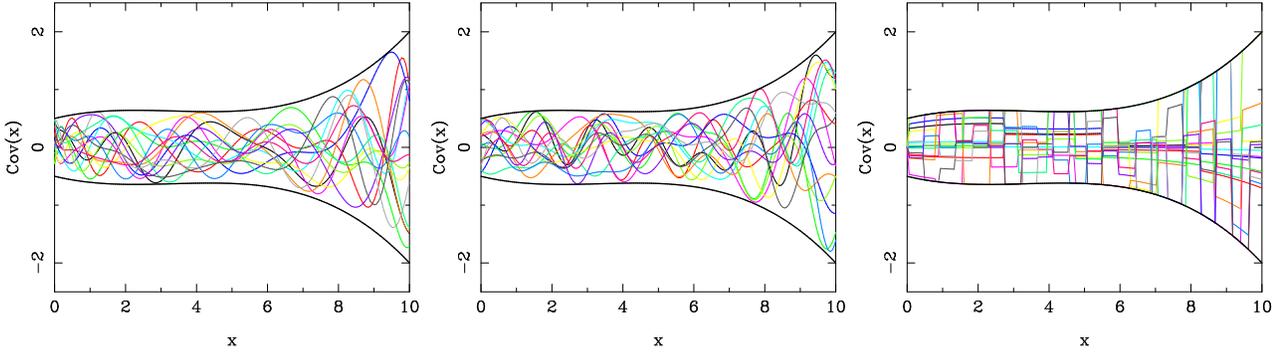

\resizebox{0.66\columnwidth}{!}{
\includegraphics{exampleBound.ps}}
\resizebox{0.66\columnwidth}{!}{
\includegraphics{exampleBoundFourier.ps}}
\resizebox{0.66\columnwidth}{!}{
\includegraphics{exampleBoundTophat.ps}}
\caption{An example of a bounded area (thick black lines), and a random sampling of $15$ 
functions shown by thin coloured (gray and black) lines. 
Every function is defined, 
using equation (\ref{E9}) so that it must fit within the bounded area. In the left panel  
we used the Chebyshev basis set, 
in the central panel we use the Fourier basis set and in the 
right panel we use the tophat basis set (binning). For all basis sets 
the maximum order is $N=15$. The functions are defined by uniformly and 
randomly sampling the coefficient space $\{a_1$, ..., $a_{15}\}$. 
This Figure is ment simply as a example of the type of 
function that can be drawn not 
as a measure of wether the method succeeds in drawing all functions. 
The success of the method in drawing all functions is shown 
in an extensive fashion in Appendix B.}
\label{exampleBound}
\end{figure*}
\\

\noindent {\bf Algorithm.} Equation (\ref{E7}) 
represents \emph{every} function, however we are only 
concerned here with defining all functions within some bounded region. 
To begin we will show how an interval $|a_n|\leq Q$ can be defined such that 
every function between $|f(x)|\leq 1$ can be drawn. For an orthonormal basis set the 
coefficients needed to draw a function $f(x)$ can be expressed as
\be
a_n=A_{nm}\int_R {\rm d}x w(x)f(x) \phi_m(x)
\ee
where $R$ is some interval over which the basis set is complete and $w(x)$ is a weight 
function upon which the basis set is complete the constant $A$
($A_{mn}$ is a diagonal) is calulated in general 
using
\be
A^{-1}_{nm}=\int_R {\rm d}x w(x)\phi_n(x)\phi_m(x)
\ee
for all the basis sets we consider $A_{nm}=A\delta_{nm}$.
Now we can write a general expression that provides a limit on $a_n$. 
Using the triangle inequality we can write 
\be
|a_n|\leq A\int_R {\rm d}x |w(x)||\phi_n(x)||f(x)|
\ee
and given that $|f(x)|\leq 1$ we have an expression for $|a_n|$
\be
\label{ann}
|a_n|\leq A\int_R {\rm d}x |w(x)||\phi_n(x)||f(x)| \leq A\int_R {\rm d}x |w(x)||\phi_n(x)| = Q.
\ee
So using equation (\ref{E7}) and limiting the coefficient values to $|a_n|\leq Q$ 
(and similarly for $b_n$) every function with the region $|f(x)|\leq 1$ can be drawn\footnote{Equation (\ref{ann}) is strictly only true for Riemann integrable functions, however essentially all bounded functions satisfy this constraint -- in particular all class $C^0$ (smooth) functions, and all step functions. Examples of the type of (very peculiar) function 
that are not Riemann integrable are Dirichlet's function and the Smith-Volterra-Cantor set.}.

We now define an arbitrary `bound function' $B(x)$ which describes a 
hard boundary (in Fig. \ref{theory_data} for example) 
where at any given point in $x$ the systematic functional form 
must lie in the region $-B(x)\leq f(x) \leq B(x)$.
Equation (\ref{E7}) is simply modified to include this arbitrary boundary 
\be 
\label{E9}
f(x; \{a_n\}, \{b_n\})=B(x)\sum^N_n a_n \psi_n(x) + b_n \phi_n(x).
\ee
Now, if the basis set is complete, $N=\infty$ and $|a_n|\leq Q$ (similar for $b_n$) 
equation (\ref{E9}) represents 
\emph{every possible} function that can be drawn within the hard bound -- 
and each function could yield a different bias. 
Note however that this 
statement says nothing about the \emph{probability} that a particular 
function will be drawn. 

An important caveat to this is that the total set of functions drawn using this algorithm 
is not bounded only some subset of the functions is. 
However a `clean' subset of functions with $|f(x)|\leq B(x)$ 
can easily be drawn by removing any function for which $|f(x)|>B(x)$ at any $x$.

In practice where $N<\infty$ the task of drawing all possible functions 
becomes a numerical/computational problem. 
For a given basis set the fundamental quantities that describe each and every function 
are the coefficients 
$a_n$ and $b_n$. The task then is to explore the coefficient 
parameter space $\{|a_n|\leq Q\}$ in an exhaustive a manner as possible. 

As a first attempt the approach taken in this article is to 
randomly and uniformly sample the coefficient space $\{a_1$, ..., $a_N\}$ for 
$|a_n|\leq Q$. 
The free parameters in this approach, given a basis set,
are the maximum order investigated $N$ and the number of 
random samples in the space $\{a_n\}$ that are chosen. 
We leave a more sophisticated Monte-Carlo formulation of this problem for future work.

When truncating the series we will be missing some highly oscillatory 
functions (for the basis sets 
considered) but we show in Appendix B that one can always make a definitive 
statement about the fraction of all functional behaviour sampled down to some scale. 
These free-form functions are regularised by the truncation of the basis set
and by the bound function.
Throughout the remainder of this Section we use the numerical 
order and function number investigated in Appendix B.
\\

\noindent {\bf Basis Sets.} Throughout we will 
principally consider three different basis sets. These are 
\begin{itemize}
\item
\emph{Chebyshev} polynomials $T_n(x)$. 
These functions form a complete basis set for $-1\leq x \leq 1$, 
and are bounded by the region $|f(x)|\leq 1$. 
To map these functions onto an arbitrary $x$-range a
variable transformation can be applied such that 
\be 
\psi_n(x)=\cos(n\arccos(x))=T_n\left(\frac{2x-x_{\rm min}-x_{\rm max}}{x_{\rm max}-x_{\rm min}}\right),
\ee
and $\phi_n(x)\equiv 0$ for all $n$. 
\item 
\emph{Fourier} series. The Fourier series is a complete basis set in the range 
$-\pi \leq x \leq \pi$ and the functions are bounded in the region $|f(x)|\leq 1$ 
for the basis set 
\ba
\psi_n(x)=\frac{1}{2}\cos\left[n\left(\frac{\pi x-\pi x_{\rm min}}{x_{\rm max}-x_{\rm min}}\right)\right]\nn
\phi_n(x)=\frac{1}{2}\sin\left[n\left(\frac{\pi x-\pi x_{\rm min}}{x_{\rm max}-x_{\rm min}}\right)\right].
\ea
The cosine or sine of a real number are $|\cos(x)|\leq 1$ and $|\sin(x)|\leq 1$. 
\item 
\emph{Tophat} functions (binning). 
This uses a tophat functional form and is meant to be analogous  
to binning the $x$-range. The maximum order in equation (\ref{E7}) $N$ 
here refers to the number of bins where for the $n^{\rm th}$ bin $\psi(x)$ is either zero or one 
depending on whether $x$ lies within the bin
\ba 
\psi_n(x)&=&
\begin{cases}
1 \quad & \forall \ (x_n-\Delta x/2) \leq x \leq (x_n+\Delta x/2)\\
0 
\end{cases}\nn
&=&H(x-\frac{\Delta x}{2})-H(x+\frac{\Delta x}{2})
\ea
where $x_n=x_{\rm min}+(n-1)\Delta x$ and $\Delta x= (x_{\rm max}-x_{\rm min})/(N-1)$ 
is the 
bin width. $\phi_n(x)\equiv 0$ for all $n$.
\end{itemize}
Table \ref{basissets} summarises some of the basis set properties including the coefficient 
intervals over which a complete (sub)set of functions with $|f(x)|\leq 1$ can be drawn.

\begin{table*}
\begin{center}
\begin{tabular}{|l|c|c|c|c|c|}
\hline
Basis Set&Basis Functions&Orthogonal Weight $w(x)$&Orthogonal Constant $A$&Interval $R$&Coefficient Interval $Q$ \\
\hline
Chebyshev&
$T_n(x)=\cos(n\arccos(x))$&
$(1-x^2)^{-\frac{1}{2}}$&
$\frac{1}{\pi}$ for $n=0$; $\frac{2}{\pi}$ for $n\not=0$&
$[-1,1]$&
$1$ for $n=0$; $\frac{4}{\pi}$ for $n\not=0$\\
Fourier&
$\frac{1}{2}$ for $n=0$, $\cos(nx)$ \& $\sin(nx)$&
$1$&
$\frac{1}{\pi}$&
$[-\pi,\pi]$&
$2$ for $n=0$; $\frac{4}{\pi}$ for $n\not=0$\\
Tophat&
H($x$-$\frac{\Delta x}{2}$)+H($x$+$\frac{\Delta x}{2}$)&
$1$&
$1$&
$[-1,1]$&
$1$\\
\hline
\end{tabular}
\caption{This table lists some constants and functions associated with the basis sets used in this article, Chebyshev, Fourier and tophat functions. Some of these constants are defined before and used in equation (\ref{ann}). The coefficient interval is such that for $|a_n|\leq Q$ all function in with $|f(x)|\leq 1$ can be drawn. 
}
\label{basissets}
\end{center}
\end{table*}


One may be concerned that there will always be some $x$ at which a boundary-touching
function cannot be drawn. For every finite 
maximum index $N$ in the sum of equation (\ref{E9}), 
one can find a value of  $x$ in $R$ such that
$f(x)$ cannot be $1$ at $x$. The key realisation that we stress here is that this is a 
actually a statement about the scale upon which the space of functions is complete. 
The larger the maximum order $N$, the smaller one can find an $\epsilon$
with $\psi_n(x\pm \epsilon)=1$ for some $n$, i.e. the concern is reduced
to an issue of resolution because if $\epsilon \ll s$ then every
function complete down to some scale $s$ can be drawn. We show this in Appendix B.

\subsection{Hard Bound}
\label{Theory Bound}

Fig. \ref{exampleBound} shows an example of a hard boundary and a 
random assortment of functions (a random, uniform, sampling of $\{a_1$, ..., $a_N\}$; and 
$\{b_1$, ..., $b_N\}$ for the Fourier basis set) 
for the Chebyshev, Fourier and tophat basis sets. 
As the order and number of functions is increased the functional 
forms begin to completely fill in the bounded area (hence ``form filling functions'').
 
It can be seen even at this stage that the tophat basis set is not efficient at 
filling the bounded area. 
This is investigated further in Appendix B where 
we show that whilst all of the basis sets considered can 
fill any desired bounded region the Chebyshev and 
Fourier basis are many orders of magnitude 
more efficient in terms of computational time than the tophat functions (binning); 
we discuss computational time in Appendix C. 
This is a 
result of the restrictive step-like nature of the functions 
that binning imposes requiring 
a high order (number of bins) to characterise particular (smooth) functional behaviour. 
In addition to being inefficient, the tophat basis is not differentiable at 
the bin boundaries and as such could be construed as being un-physical, although 
in some special 
circumstances (e.g. photometric redshifts where a filter may have a tophat
function in wavelength) a tophat basis set may be needed.

The key feature of the hard bound is that within the boundary all functions are given 
equal weight i.e. the probability that any given function 
is the `true' systematic functional 
form is the same for all functions. 
Out of the full range of possible biases given a hard boundary, there should exist 
a \emph{maximum bias} -- because the space of functions and hence the range of 
biases is limited. We show that this is the case in Appendix D.
The quantity of interest is therefore 
this maximum bias that is allowed by the functions that can be drawn within the hard 
boundary. In the data boundary case, Section \ref{Data Bound}, 
there is a maximum bias for each subset of functions that give the same weight (with 
respect to the data).

One intuitively 
expects that the systematic function that introduces the largest 
bias should be the one that 
most closely matches the signal term containing each parameter. 
In the case where there are 
degeneracies between parameters the worst function is some 
combination of the signals 
sensitivity to the parameters. 
Here we will use the simple example hard bound from Section 
\ref{A Simple Toy Model} to 
demonstrate that in this case a maximum bias exists, and that this maximum 
is stable with respect to basis set. 

\begin{figure}
\resizebox{84mm}{!}{
\includegraphics{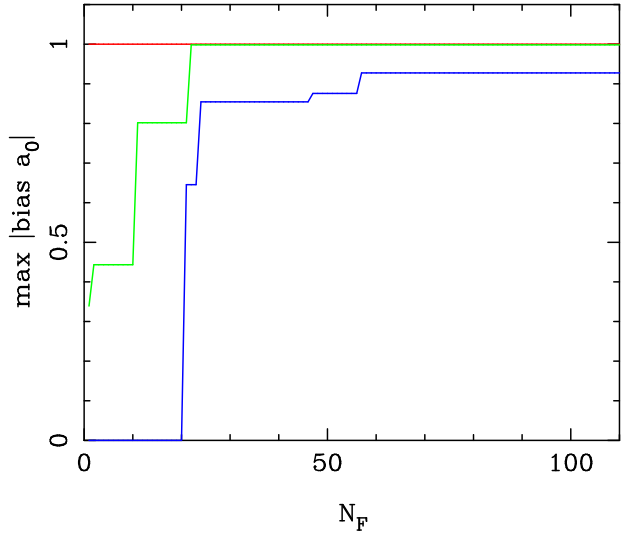}}
\caption{Using the simple example from Section \ref{A Simple Toy Model}. 
The maximum bias on $a_0$ as a function 
of the number of realisations of the systematic basis set's coefficients. Red 
(dark gray, upper line) 
shows the maximum bias found using the Chebyshev basis set, 
green (light, middle line) gray shows the maximum 
bias for the Fourier basis set and blue (darkest gray, lower line) for the tophat basis set.}
\label{bias_vs_NF}
\end{figure}

Fig. \ref{bias_vs_NF} shows the maximum bias on $a_0$ as a 
function of the number of random 
realisations of the coefficient space for the Chebyshev, Fourier and tophat systematic 
basis sets, for all sets we consider a maximum order of $100$. 
It can be seen that as soon as the ``worst function'' is found the maximum bias becomes 
constant and stable for the Chebyshev and Fourier basis sets. In this case the worst function
is simply $C^{\rm sys}(x)=x$ since this is the function 
that affects the signal 
($C^{\rm sig}(x)=a_0+a_1x+{\rm constant}$) the most through the effect
on $a_1$ (the worst function is a linear combination of the response
of each individual parameter in the marginalised case, see Appendix D). 
There is an exact degeneracy since a function of the form $-f(x)$ will cause a bias of equal 
magnitude but opposite sign to $f(x)$, so $C^{\rm sys}(x)=-x$ in this case will also 
cause the same absolute bias (see Appendix D). For the top hat basis set 
it is very unlikely to find this particular worst function so the bias does not find the maximum 
even after $100$ realisations.
The Chebyshev basis set finds this function after only a few realisations 
since $\psi_1(x)=T_1(x)=x$ is one of the basis functions of the Chebyshev expansion. 

Fig. \ref{bias_vs_NF} is meant simply as an example of the type of convergence test that 
could be performed given a realistic application. In Appendix B we outline a diagnostic 
mechanism that can gauge whether a bounded area has been completely 
filled with functions.

\subsection{Data Bound}
\label{Data Bound}
If the systematic has been partially measured using data then 
there is no hard boundary within which all functions must lie, and every 
function is still allowed to some degree. The issue which must be addressed in this case is 
how each function should be weighted given the data available -- a function is more 
likely to be the `true' systematic signal if it is a good fit to the systematic data. 

For the case of a data bound we propose a similar approach to the hard bound. We want to 
explore all possible functions, to do this we express an 
arbitrary function using an expansion as in 
equation (\ref{E9}). By choosing a complete basis set (e.g. Chebyshev, Fourier)
the full space of functions can be explored by exhaustively exploring the 
space of coefficients, as described in Section \ref{Theory Bound}.

In the following we will 
concatenate the two basis sets $\{a_n\}$ and $\{b_n\}$ for clarity, however note that 
for the Fourier basis set two sets of coefficients are needed.

The critical difference in this case is that for each function $f(x;\{a_n\})$ 
that is drawn we can assign a weight 
$W(\{a_n\})$ using the $\chi^2$ statistic
\be 
\label{E10}
\chi^2=\sum_x\frac{[f(x; \{a_n\})-\widetilde{C^{\rm sys}}]^2}{\sigma^2_{C^{\rm sys}}}.
\ee
This quantifies how good a fit the function (the values of $a_n$ and $b_n$, 
and some basis set) is to the residual systematic data. 

However since we aim to exhaustively try 
every function (complete down to some scale) there will 
exist some functions that fit \emph{exactly} through the data points. 
This causes a problem when using
the $\chi^2$ statistic as a weight for such functions will have a
$\chi^2\equiv 0$, and  
hence be given a probability $P=1$ that such a function is likely, but such a 
conclusion has not taken account of the error bars on the systematic data. 

In this simple example we have subtracted the mean of the systematic 
signal so data should be scattered about $C(x)=0$. In the case of
Gaussian distributed  
data the scatter of the data points should be proportional to the error bar 
on each data point. 
If the observation could be repeated then the data points would be
scattered in the same  
statistical manner about $C(x)=0$ but have different actual values. 
This is very similar to the 
familiar sample variance; in cosmology we are used to a special kind of sample 
variance that we call cosmic variance in which the likelihood of the data, given 
only one realisation of our Universe, must be taken into account. 

To take into account this sample variance effect we must consider the
likelihood of the  
residual data. In Appendix E we show how for 
Gaussian distributed data the probability of 
a function $f(x;\{a_n\})$, given some set of observations, 
can be written as a sum over $x$, where at each point there is some 
data with an error bar $\sigma_{C^{\rm sys}}$ as
\ba
\label{E11}
\ln p[f(\{a_n\})]\propto &-&\sum_x\left[\frac{f(x;
    \{a_n\})^2}{4\sigma_{C^{\rm sys}}^2}\right]\nn 
&+&\sum_x\ln(\sigma_{C^{\rm sys}}\sqrt{\pi})
\ea
where we have translated the notation of equation (\ref{B5}) to reflect that of 
equation (\ref{E10}). For any given set of data the second term in
equation (\ref{E11}) is a  
benign additive constant so the weight that we assign each function is 
\be
\label{E12}
W(\{a_n\})=\sum_x\left[\frac{f(x; \{a_n\})^2}{4\sigma_{C^{\rm sys}}^2}\right].
\ee
We stress that this is for Gaussian data with a known mean of zero
such a function is uniquely described by the variance. 

In general data
may not be exactly centered at zero or be Gaussian distributed.
As an alternative to the analytic 
procedure of marginalising over the data one could create Monte-Carlo 
realisations. For each realisation one could measure the $\chi^2$ of the fit of a 
given function to the data from equation (\ref{E10}) and assign a weight as the 
average over all realisations 
\be
\label{E13}
W(\{a_n\})=\frac{1}{2}\left\langle \sum_x\frac{[f(x;
    \{a_n\})-\widetilde{C^{\rm sys}}]^2}{\sigma^2_{C^{\rm
      sys}}}\right\rangle_{\rm realisations} 
\ee
where the extra factor of $1/2$ converts the average $\chi^2$ to a log-likelihood similar 
to equation (\ref{E12}).

The bound function in equation (\ref{E9}) represents a hard 
prior in this case. One should choose a bound function is much larger than the scatter in the 
systematic data $B(x)\gg \sigma_{\rm sys}(x)$. 
Any functions that deviate from the data 
by a large amount will be down-weighted by the poor fit even though they 
may yield a large bias, so as long as the boundary within which functions are considered is 
$B(x)> 3\sigma_{\rm sys}(x)$ away from the data then any results should be robust. 

For each function $f(x; \{a_n\})$ one now has an 
associated bias and a weight. Now consider a 
particular bias in some parameter: there exists a set of systematic 
functions that could yield this same bias and from that set there must exist \emph{a} 
function which is the best fit to the data. If each best-fitting function for 
every bias can be 
found then one is left with a robust weight for each bias and 
a measure of the relative probability allowed by the data
\be 
\label{like1}
p(b_i)\propto \exp(-\min[W(b_i; \{a_n\})]).
\ee
$\min[W(b_i; \{a_n\})]$ is the minimum 
weight (equation \ref{E12}) from the space of functions defined 
by the coefficients $\{a_n\}$ that yields the bias $b_i$.

Another way of putting 
this is that for a given weight there exists a maximum and minimum bias. 
We show in Appendix D that, 
within the Fisher matrix approximation of equation (\ref{biase}), 
that there does 
indeed exist a maximum and minimum bias for each weight. 

\begin{figure}
  \resizebox{40mm}{!}{
    \rotatebox{0}{
      \includegraphics[clip=true]{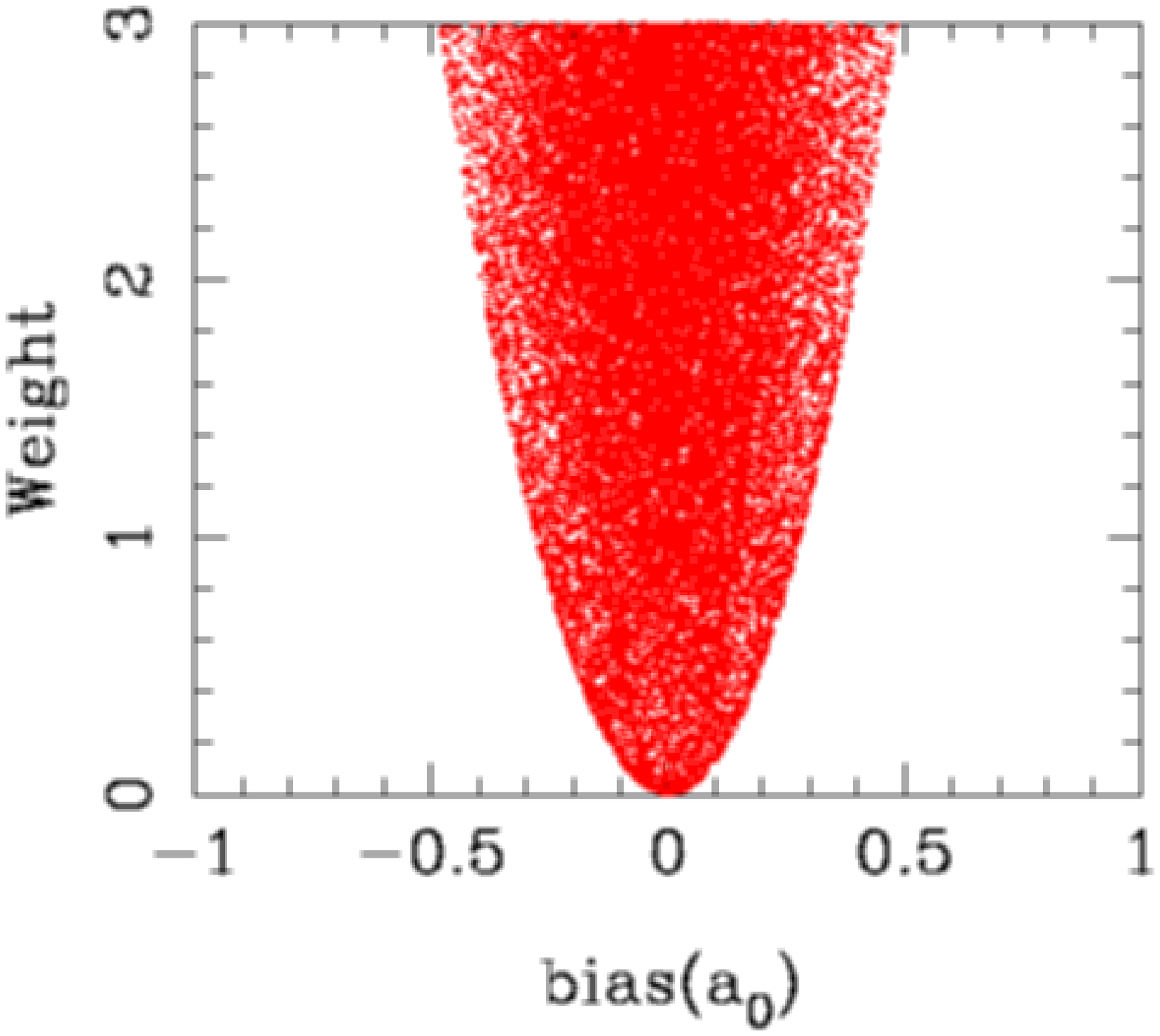}}}
  \resizebox{40mm}{!}{
    \rotatebox{0}{
      \includegraphics[clip=true]{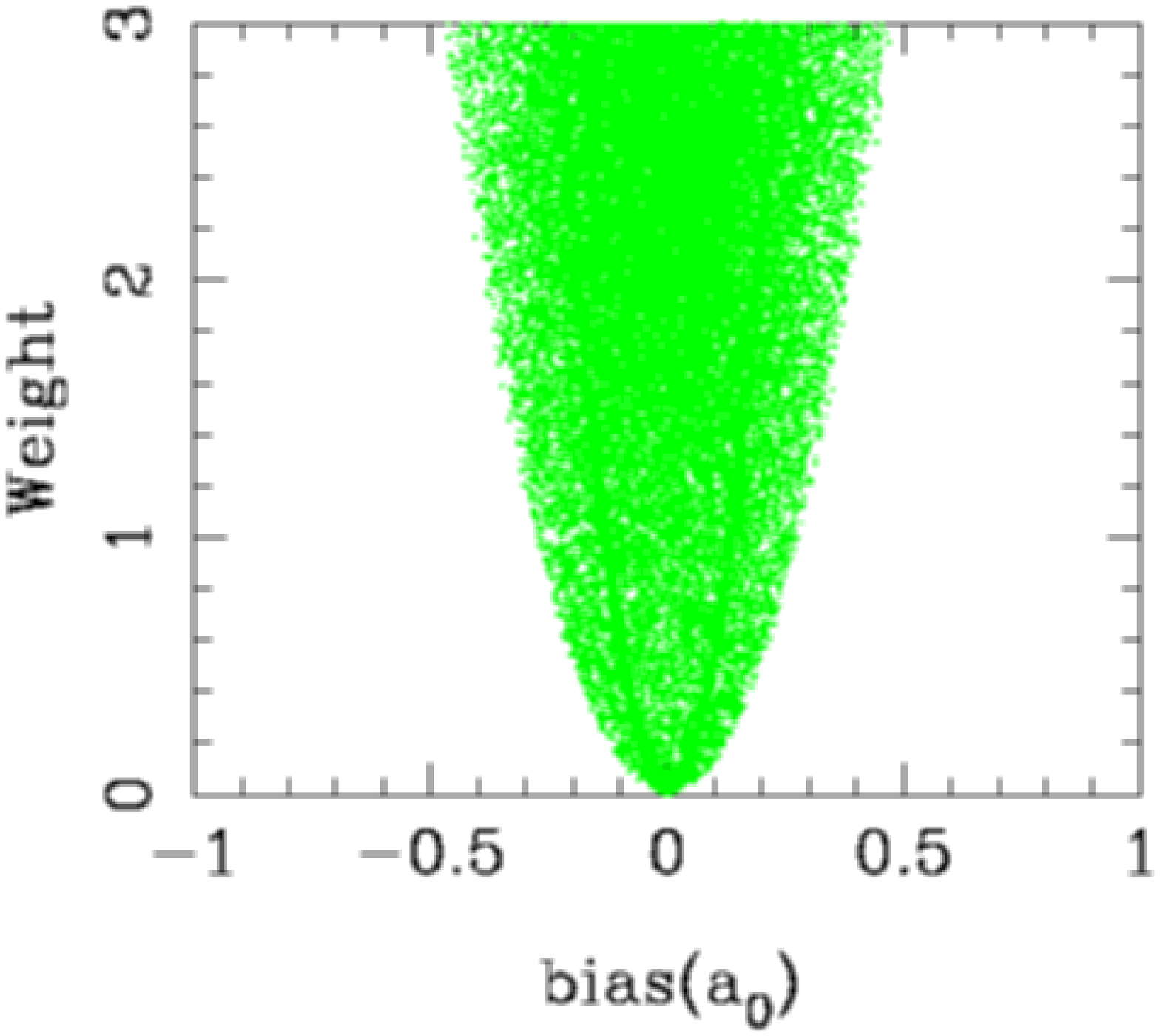}}}\\
  \resizebox{40mm}{!}{
    \rotatebox{0}{
      \includegraphics[clip=true]{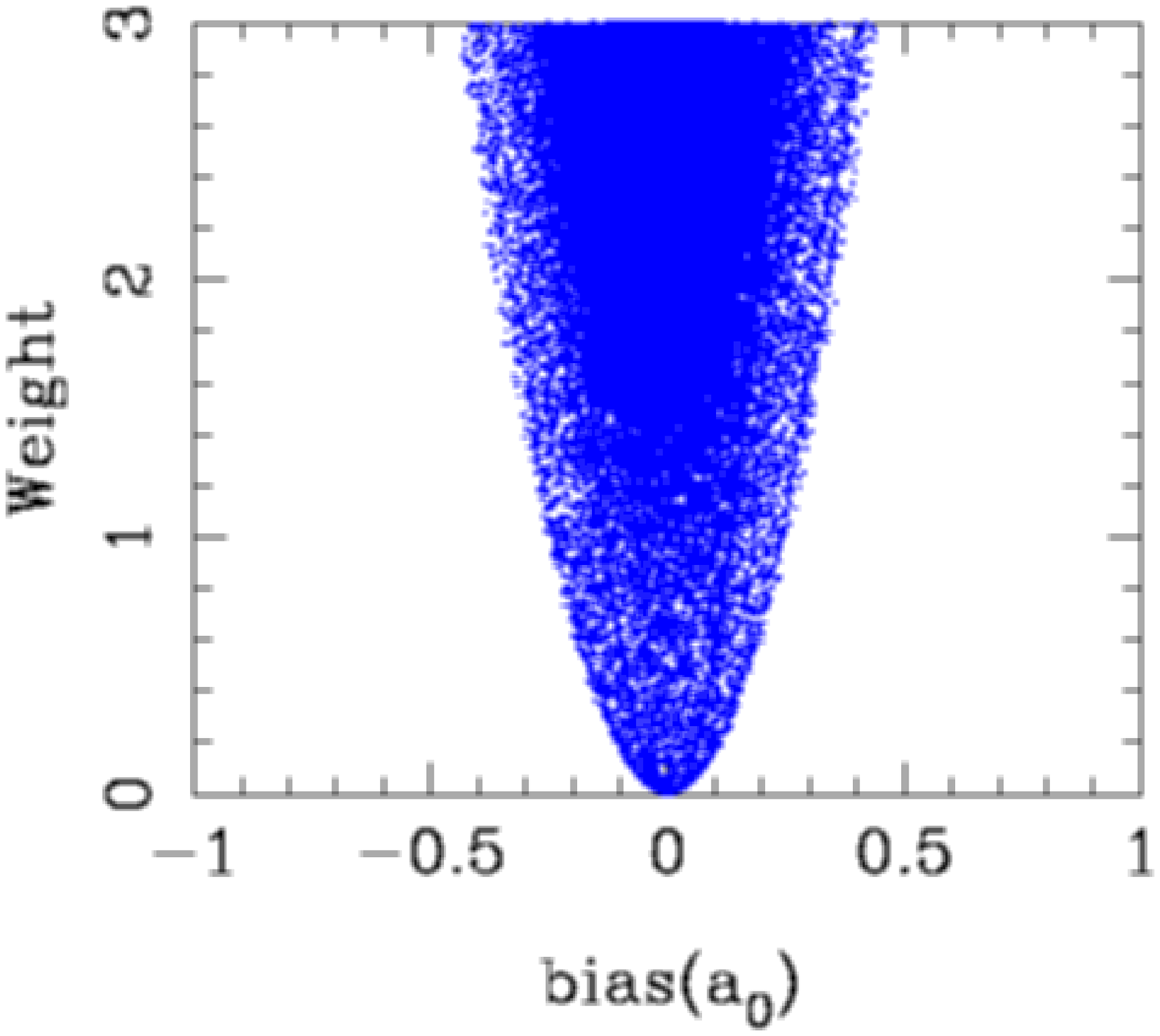}}}
  \resizebox{40mm}{!}{
    \rotatebox{0}{
      \includegraphics[clip=true]{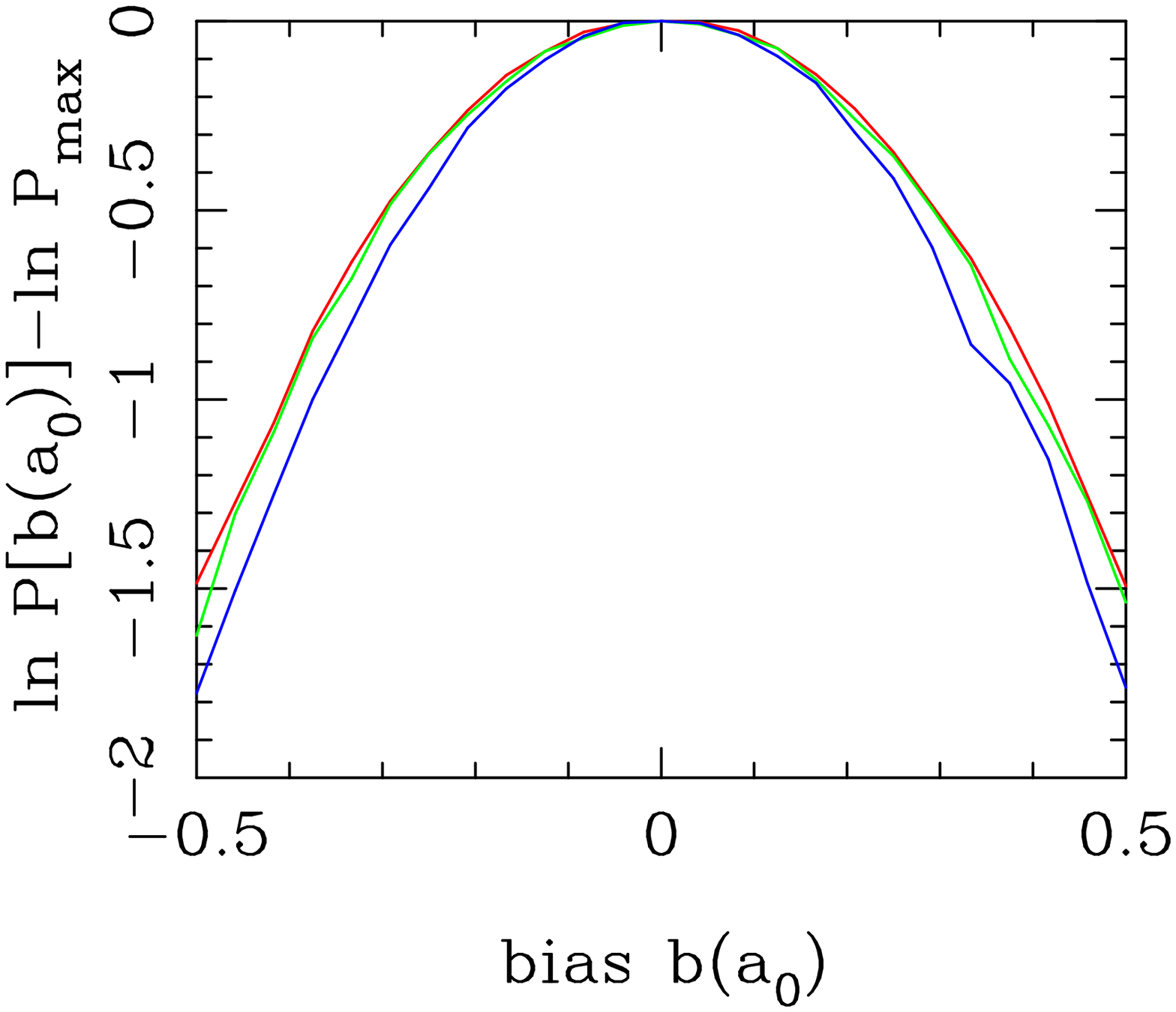}}}
    \caption{The three scatter plots show the bias in the parameter $a_0$ caused by 
      fitting functions through the data bounded systematic shown in Fig \ref{theory_data} 
      against the weight given to the function with respect to the systematic data points given
      in equation (\ref{E12}). 
      Each point represents a function. The colours correspond to the different basis sets 
      considered blue (darkest gray, bottom left) is tophat functions, 
      red (lighter gray, top left) is Chebyshev 
      functions and green (lightest gray, top right) is Fourier functions. Also shown is 
      the likelihood 
      of the bias in $a_0$ for each basis set, found by measuring the minimum weight for a 
      particular bias -- the lowest extent of the scattered points in the other plots -- 
      and using equation (\ref{like1}).}
    \label{ChiExample}
\end{figure}

Fig. \ref{ChiExample} shows the weight (equation \ref{E12}) 
for each function drawn from the Chebyshev, Fourier 
and tophat basis sets against the bias induced by the function using the data bound 
toy model of Fig. \ref{theory_data}. It can be seen from this Figure that for any given bias 
there exists a minimum weight that can be achieved by the functions giving that bias. 
In the right bottom panel of Fig. \ref{ChiExample} we
have found the minimum weight for each bias and converted this into a likelihood using 
equation (\ref{like1}) showing that this procedure is robust to the choice of basis set used
(though the tophat basis set is far less efficient than Chebyshev and Fourier, see 
Appendix B for more details). 

\begin{figure}
\resizebox{84mm}{!}{
\includegraphics{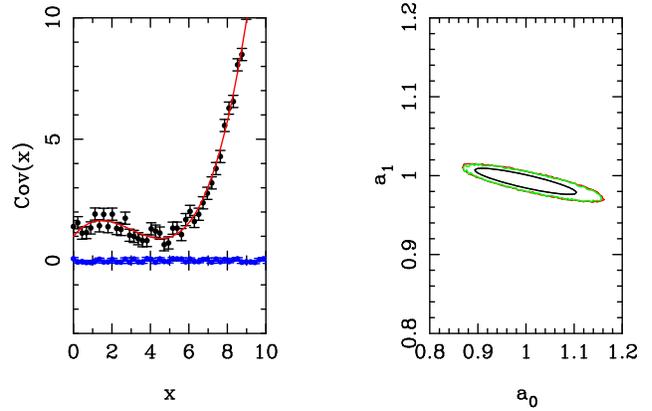}}
\caption{The left hand panel shows the modified simple example, 
see Section \ref{A Simple Toy Model}, where we have made a more 
tightly constrained Gaussian residual systematic with $\sigma_{C^{\rm sys}}=0.05$ for clarity, 
this can be compared with the right hand panel in Fig. \ref{theory_data}. The right hand 
panel shows the $1-\sigma$ statistical constraints on $a_0$ and $a_1$ (black solid line). 
In addition we show the \emph{systematic} probability contours for the residual data bound in 
the left hand panel. Shown are the $1-\sigma$ systematic contours using both a 
Chebyshev (red, dark gray line) and a Fourier (green, light gray line) basis set for 
functional form filling.}
\label{simple_contour}
\end{figure}

One can of course extend the formalism introduced here 
to an arbitrary number of dimensions. To extend this to two dimensions we have 
modified the simple example to have a much smaller data bound, since (as can be seen in Fig. 
\ref{ChiExample}) the fiducial scatter in the toy model systematic data 
introduces large biases in the measured parameters. 
Fig. \ref{simple_contour} shows the \emph{joint} $1-\sigma$ statistical 
constraint on $a_0$ and $a_1$ for the mock data shown. 
On this same plot we show the systematic 
$1-\sigma$ contours from the residual systematic -- within the systematic contour there is a 
$\geq 68\%$ probability that the statistical maximum likelihood is biased. 

We again show results for two different 
basis sets and show that the systematic contours are again \emph{independent} of the basis set 
chosen. We have not used the tophat basis because it is not an efficient basis set (in terms of 
computational time, see Appendix B) to use even for a one dimensional analysis. 

In a real application one would hope to show such plots with the 
systematic contours lying within
the statistical ones, but for illustrative purposes we have shown a dominant systematic here.
A hard bound in this case would represent a tophat contour in probability where, 
within the contour, the probability equals unity and outside the contour 
the probability is zero.

In the marginalisation approach one could also draw two sets of contours, but both
would be statistical: one that has no systematic and the other in which a systematic 
has been included. Fig. \ref{simple_contour} represents one of the key recommendations 
of this article, that in
future we must not only show statistical contours for cosmological parameters 
but also systematic 
probability contours. Here we have shown one way of obtaining a robust estimate of such 
systematic probability contours.  

\subsection{Marginalisation vs. Bias}
\label{Marginalisation_Bias}
The bias and the marginal error are inter-related via the mean 
square error (MSE) which is defined, for a parameter $a_i$, as
\be
\label{MSE}
{\rm MSE}=\sigma^2(a_i)+b^2(a_i)
\ee
where $\sigma(a_i)$ is the marginal error on $a_i$ and $b(a_i)$ is the bias. So both 
marginalisation over systematic parameters and functional form filling increase the MSE 
through the marginal error and bias respectively. 

In Appendix A we show that the marginalising approach and the functional form 
filling approach are different however there is a subtelty. \emph{If}
the parameterisation and order are the same (i.e. a truncated basis
set is used) then the MSE recovered from marginalising or considering
the bias is the same. The conclusions from Appendix A are
\begin{itemize}
\item 
\emph{If} the functional form of the systematic is known then the
degradation in the MSE as a result of marginalising or functional form
filling (bias) is the same.
\item 
If the functional form of the systematic is unknown then the MSE from
marginalising will tend to underestimate the true systematic error and
is in general not equal to the MSE from functional form filling.
\item 
Given that the marginalising necessarily truncates the basis set there
are always some functions that marginalising cannot assess. 
\end{itemize}

\begin{figure}
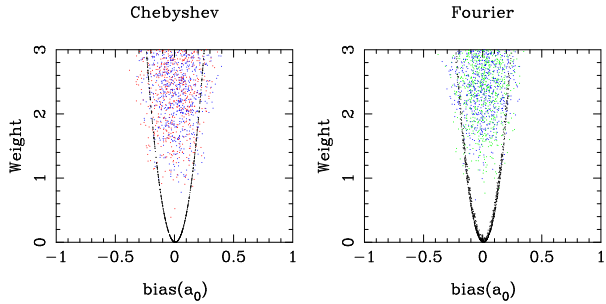

  \resizebox{40mm}{!}{
    \rotatebox{0}{
      \includegraphics[clip=true]{margbias_cheb.ps}}}
  \resizebox{40mm}{!}{
    \rotatebox{0}{
      \includegraphics[clip=true]{margbias_four.ps}}}\\
    \caption{The weight (fit to the systematic data) against the bias
      in the parameter $a_0$ for the toy data bounded systematic shown
      in Fig \ref{theory_data}. We trucate the Chebyshev and Fourier
      basis sets at $N=1$ (black dots) $N=10$ (red, green -- light
      gray dots) 
      and 
      $N=50$ (blue, dark gray dots). For each order we make $500$ 
      realisations of
      the basis set.}
    \label{marg_vs_bias}
\end{figure}

Given that the MSE, given a particular functional subset is the same for
bias and marginalisation the key difference between marginalisation
and functional form filling is that in marginalisation case the
functional space is truncated by the choice of parameterisation and
ultimately by the number of data points. 

Looking back at Fig. \ref{ChiExample} the functional form filling
technique fills out the Weight--MSE (bias) bounded region by sampling 
every function down to some scale. To
illustrate the way in which marginalisation cannot sample the full
space of functions we have reproduced these plots but using various
truncated basis sets. Note that the MSE from the bias and
marginalisation is the same given a parameterisation as shown in
Appendix A -- marginalising is like functional form filling but with a
very restricted class of function.

Fig. \ref{marg_vs_bias} shows the weight (fit to data) vs. bias using
the simple toy model for
the Chebyshev and Fourier truncated basis sets for $500$ realisations
of each truncated set. We truncate the
expansion at $N=1$, $10$ and $50$. It can be seen that if $N=1$ is used
the space of functions is very restricted, and as the order is
increased the space of functions increases. These results are also in
resonance with Appedix B. If one were marginalising a choice of
parameterisation (basis and order) would have to made in which case
the MSE conclusion would be dependent on this choice. 
\\

\noindent {\bf Discussion}

We take this opportunity to discuss the difference between marginalising 
and functional form filling. Functional form filling is not the same 
as marginalising over a very large parameter set on a qualtitative level: 
we are not finding the best fit values of the 
parameters but rather using each parameter combination simply as a 
prescription that yields a 
particular bias. As the freedom in the functions increases (more parameters) 
the functional form filling approaches a stable regime in the results it gives. 
Functional form filling yields the probability 
that the maximum likelihood value is biased 
by some amount, whereas marginalisation yields a probability that the cosmological 
parameters take 
some value jointly with some values of the nuisance parameters.

One could marginalise over a very large number of parameters but 
in this case the joint covariance matrix 
will at some point necessarily become computationally singular as parameters
are introduced can cannot be constrained by the data. 
One way to consistently include more parameters than data points $N$ is to 
add a prior such that any extra parameters were constrained 
$P(A|D) = \int dB P(A,B|D)P(B)$ where $A=N$ ansd $(A+B)>N$; but this
requires adding the prior on $P(B)$ which requires justification. 

In the lack of any external prior  
the number of parameters that can be used in marginalising 
is necessarily
truncated at a low order since once the number of free 
parameters becomes larger than the number of data points (for the data bound) 
the parameter fitting methodology becomes ill-posed (e.g. Sivia, 1996). 
In contrast the functional form 
filling technique could use an infinite basis set expansion 
(only computational, and physical, resources prohibit this) 
since we use the basis set simply as \emph{a prescription for drawing functions} 
through the systematic -- one could even 
draw these by hand if you were sure that you could draw a 
complete set of functions. 

One may be concerned that information is being ignored or disgarded. 
Quite the opposite from discarding information present in data we  
assign a weight to every possible function with respect to the data, in this sense 
we throw nothing away -- every function has a weight and a bias. In constrast 
when marginalising the functional space assessed is limited by the number of data
points and as such a very large class of functions are never  
considered -- if a basis set is truncated these are usually highly
oscillatory functions.

The bias formalism becomes preferable if there is 
uncertainty over what the functional form is (which is most, 
if not all of the time). 

In Appendix A we show that for the same basis set and prior functional
form filling and marginalisation produce the same results in a
mean-square-error sense. Hence in
the limit of marginalising over 
functions, with suitable priors, the two approaches should produce the same results -- in
this article we focus on bias functional form filling. 

\subsection{Summary}

We have presented an alogorithm by which any function within a bounded region 
can be drawn this is summarised as 
\begin{itemize}
\item
Chose a complete orthogonal basis set. We recommend Chebyshev polynomials because 
of their ease of calculation and computational efficiency (see Appendix B).
\item 
Use equation (\ref{ann}) to calculate the interval $a_n \in [-Q,Q]$ 
over which the coefficient must be sampled.
\item 
Define the bounded region $B(x)$ within which functions must be drawn. 
\item 
Define a scale $\Delta x$ upon which the functional space must be complete. 
\item 
Randomly and uniformly sample from the space of coefficients using a maximum 
order $N$ and number of realisations such that the functional space is fully 
filled -- the diagnostic tools from Appendix B can be used to guage the level 
of completness.
\end{itemize}

By defining the hard and data bounds we have now presented all 
the tools needed to correctly 
assess a systematic given some prior knowledge of the magnitude of the effect, 
either an external data set or some theoretical knowledge. In the 
hard boundary case every function is given equal weight and as such a maximum bias 
should exist. In the data boundary case a probability can be assigned
to each bias. 

We emphasise here that this method requires there to be at least
\emph{some} information at every data point associated with the
signal -- either a hard boundary or some systematic data. If there
were no constraint the ranges of biases could become unbounded.

Our proposal is that when measuring some cosmological 
parameters these techniques can be used to augment the statistical marginal 
error contours: some 
cosmological parameters are measured and about their maximum likelihood point 
are drawn some 
marginalised statistical error contours and in addition:
\begin{itemize} 
\item 
If a theory or simulation provides a 
hard boundary to some systematic then the maximum bias will define a 
systematic contour that can be drawn in within which the 
maximum likelihood could be biased. 
\item
If some data is provided that measures the systematic then a further set of systematic contours can be drawn which will show the 
probability that the maximum likelihood point is biased by any particular amount. 
\end{itemize}
The goal for any experiment is to control systematic effects to such a degree that 
any systematic contours drawn are smaller than the statistical contours.

We will use the functional form filling approach in the next Section to place 
requirements on weak lensing systematics. For general conclusions please skip to 
Section \ref{Conclusion}.

\section{An Application to Cosmic Shear Systematics}
\label{An Application to Cosmic Shear Systematics}

We will now use the functional form filling approach to address shape measurement systematics  
in cosmic shear (due to methods Heymans et al, 2006, Massey et al., 2007; or PSF inaccuracies 
Paulin-Henriksson et al., 2008; Hoekstra, 2004). 
This Section represents an extension of the work of 
Amara \& Refregier (2007b) where certain particular functional forms were investigated. 
Here we extend the analysis to include all functional behaviour to fully 
address the impact of the systematic.
For a thorough exposition of cosmic shear  
we encourage the reader to refer to these 
extensive and recent 
reviews and websites (Munshi et al., 2006; Bartelmann \& Schneider, 2001; Wittman, 2002; 
Refregier, 2003; {\tt http://www.gravitationallensing.net}). 
The source code related to the 
work in this Section is released through {\tt http://www.icosmo.org}, 
see Appendix F for details. 

\subsection{Background}
Cosmic shear tomography uses both the redshift of a galaxy and the gravitational lensing 
distortion, shear, to 
constrain cosmological parameters. The observable in this case is the lensing power spectrum 
as a function as redshift and scale $C_{\ell}(z)$. 
Since we have redshift information the galaxy population is split into redshift bins 
where each bin has its own auto-correlation function and the cross-correlations 
between bins are also be taken. 

Throughout this Section we will use a fiducial cosmology of $\Omega_m=0.3$, $\Omega_{DE}=0.7$, 
 $\Omega_B=0.0445$, $h=0.7$, $w_0=-0.95$, $w_a=0.0$, $\sigma_8=0.9$, $n_s=1.0$ where we 
parameterised the 
dark energy equation of state using $w(z)=w_0+w_a(1-a)$ 
(Chevallier \& Polarski, 2001; Linder, 2003) 
and included the spectral index $n_s$, we consider non-flat models throughout where 
$\Omega_k=1-\Omega_m-\Omega_{DE}$. We assume a weak lensing survey (similar to the 
DUNE/Euclid proposal, Refregier et al., 2008a) which has an Area$=20,000$ square 
degrees, a median redshift of $z_m=0.8$ (using the $n(z)$ given in Amara \& Refregier, 2007b) 
with a 
surface number density of $40$ galaxies per square arcminute. 
We also assume a photometric redshift 
error of $\sigma_z(z)=0.03(1+z)$ and split the redshift range into $10$ tomographic bins. 

The observed lensing power spectrum can be written as a sum of signal, 
systematic and noise terms 
\be 
C^{\rm obs}_{\ell}=C^{\rm lens}_{\ell}+C^{\rm sys}_{\ell}+C^{\rm noise}_{\ell}
\ee
so that an estimator of the observed lensing power spectrum can be written by subtracting 
the shot noise term $C^{\rm noise}_{\ell}$ 
\be
\label{WL1}
\widehat{C^{\rm lens}_{\ell}}=C^{\rm lens}_{\ell}+C^{\rm sys}_{\ell}
\ee
where $C^{\rm lens}_{\ell}$ is the underlying true lensing power spectrum, 
dependant on cosmology,  
which we want to measure and $C^{\rm sys}_{\ell}$ is some residual systematic. 
The error on this estimator can be written as 
\be
\label{WL2}
\Delta C_{\ell}=\sqrt{\frac{1}{(2\ell+1)f_{\rm sky}}}[C^{\rm lens}_{\ell}+C^{\rm sys}_{\ell}+C^{\rm noise}_{\ell}]
\ee
note that this is the error on the observed signal, not the observed signal 
itself, so that as the systematic increases the error on the observed $C_{\ell}$ 
increases.
The Fisher matrix and bias are then defined in the usual way 
(equations \ref{fishe} and \ref{biase}) where $\sigma_C=\Delta C_{\ell}$. 
\begin{table}
\begin{center}
\begin{tabular}{|l|c|}
\hline
Parameter&Marginal Error\\
\hline
 $\Omega_m$&$0.006$\\
 $\Omega_{DE}$ &$0.036$\\
 $\Omega_B$&$0.015$\\
 $h$&$0.086$\\
 $w_0$& $0.046$\\
 $w_a$& $0.152$\\
 $\sigma_8$&$0.009$\\
 $n_s$&$0.019$\\
\hline
\end{tabular}
\caption{The marginal error on each cosmological parameter for the 
fiducial weak lensing survey with no residual systematic. Note 
that no priors have been added on any parameter.}
\label{fiderrors}
\end{center}
\end{table}
Table \ref{fiderrors} shows the expected marginal errors on the cosmological 
parameters using the 
fiducial survey calculated using the Fisher matrix formalism, 
note that no prior has been added to these results, they are for lensing alone.

The additive systematic which we consider here is a special kind that has the same shape 
as the lensing power spectrum 
but where it is multiplied by some unknown systematic function (sometimes referred to as a 
multiplicative systematic) $A_1$ (we use the notation of Amara \&
Refregier, 2008b)
\be
\label{WL3}
C^{{\rm sys,}ij}_{\ell}=A_1 C^{ij}_{\ell}
\ee
where $ij$ means the correlation between redshift bins $i$ and $j$, an auto-correlation 
is where $i=j$. The particular form of multiplicative bias we 
consider is that which causes the 
true shear, as function of angle and redshift $\gamma^{\rm lens}(\theta,z)$, to be incorrectly 
determined such that the residual systematic shear $\gamma^{\rm sys}(\theta,z)$ 
is related to the true shear by some function $m(z)$ 
\be
\label{WL4}
\gamma^{\rm sys}(\theta,z)=m(z)\gamma^{\rm lens}(\theta,z).
\ee
So that the systematic given in 
equation (\ref{WL3}) can be written $C^{{\rm sys,}ij}=[m(z_i)+m(z_j)]C^{ij}_{\ell}$. 
This expression from  Amara \& Refregier (2007b) actually assigns an $m(z)$ 
to each redshift slice where each $m(z)$ is the bin weighted average. 
A more accurate approach is to include $m(z)$ in the integrand of the lensing 
kernel ($\xi_{+/-}$ in Appendix A of Amara \& Refregier, 2007b). We compared 
results when the bin-weighted average was used against the correct integral 
method and found exact agreement, this is because the bins are narrow and 
the functional variation on sub-bin width scales is small. 

We note that Huterer et al. (2006) have used Chebyshev polynomials in weak lensing systematic 
analysis, although they marginalise over a systematic parameterised using 
Chebyshev polynomials 
and then investigate the biases introduced if the most likely value of the estimated 
coefficients were incorrect. Here we are not measuring the Chebyshev coefficients but rather 
using this basis set to draw every possible function and determine the bias introduced by the 
function itself not some misestimation of any particular coefficient.
The approach outlined in this article 
also has a resonance with the work of Bernstien (2008) in which flexible 
basis sets are used to address systematic quantities, although this is done within the 
self-calibration (marginalising) framework, and their general thesis is that of 
tuning the models that are marginalised over. Bernstien (2008) notes that an 
approach such as the one outlined in this article, using the formalism of 
Amara \& Refregier (2007b), would be desirable in certain situations. 

\subsection{Functional Form Filling for m(z)}

We will now place constraints on the function $m(z)$ such that 
that the measurement of the cosmological parameters are robust. We first define some boundary 
(bound function) on $m(z)$ such that any systematic function must lie within the 
bounded region. We parameterise the hard boundary (Section \ref{Theory Bound}) using 
\be 
\label{WL5}
|m(z)|=m_0(1+z)^{\beta},
\ee
we want to know what values of $m_0$ and $\beta$ are sufficient to ensure that the bias on 
cosmological parameters $b(\theta_i)$ are less than their statistical error 
$\sigma(\theta_i)$; $|b(\theta_i)/\sigma(\theta_i)|\leq 1$. We stress that this is a 
parameterisation of the \emph{boundary} within which a function must lie, not the 
function itself. 

To fill the bounded area with every functional form, complete down to the scale of 
$\Delta z=0.2$, 
we use Chebyshev polynomials with a maximum order of $N=35$ and investigate 
$N_F=10^4$ realisations 
of the basis set, using the formalism outlined in Section \ref{Functional Form Filling}. 
We do not expect any of the cosmological parameters to introduce features into the 
lensing power spectrum on scales $\Delta z < 0.2$ so this should be sufficient.
\begin{figure}
\resizebox{84mm}{!}{
\includegraphics[clip=true]{WLnumeric.ps}}
\resizebox{84mm}{!}{
  \rotatebox{0}{
    \includegraphics[clip=true]{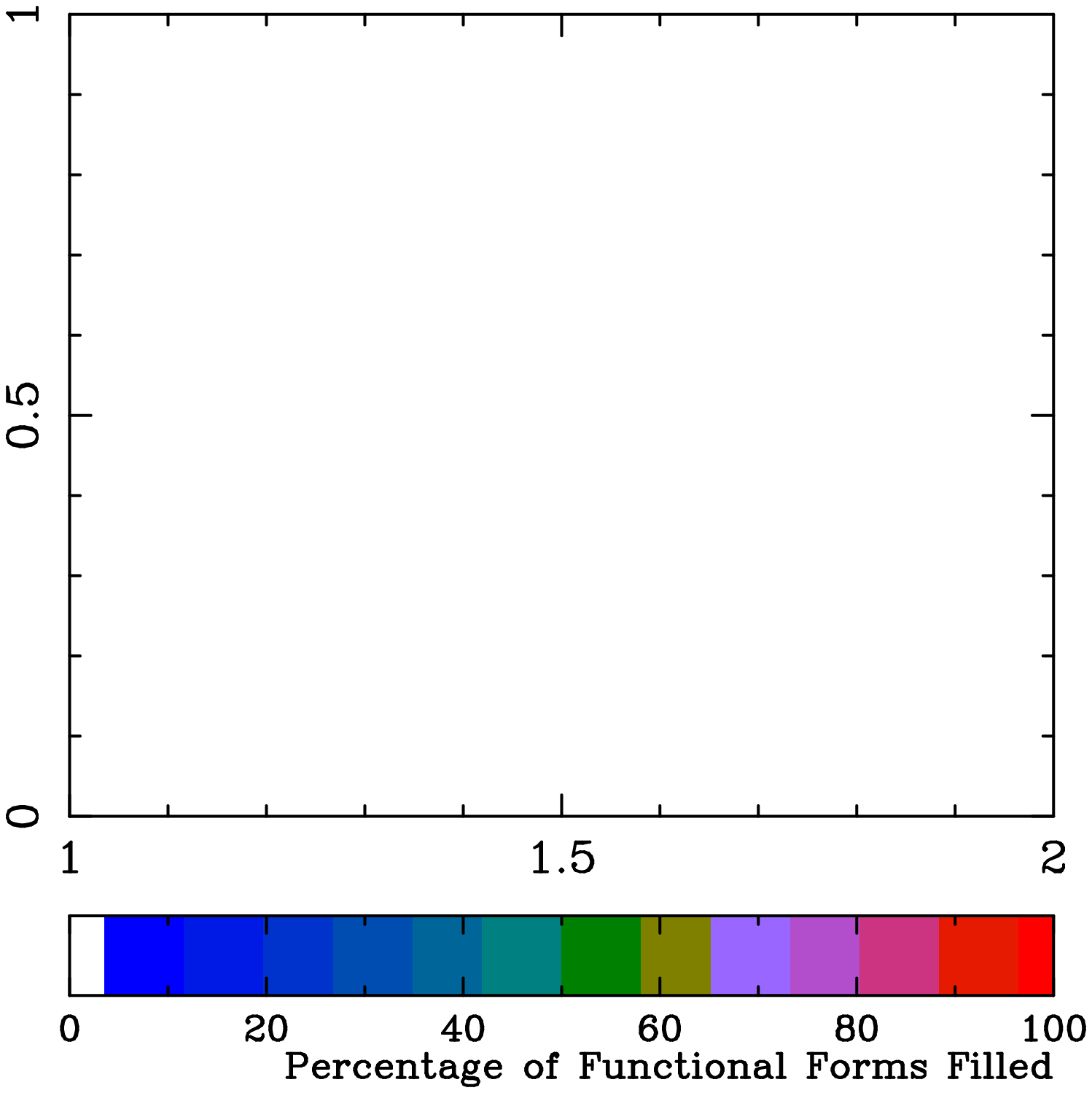}}}
\caption{This plot shows the bounded area defined using equation (\ref{WL5}) 
with $m_0=1\times 10^{-3}$ and $\beta=1.0$, shown by the solid black lines. 
This area is then pixelated and the 
functional behaviour that can occur at each pixel is measured. The colours correspond to the 
percentage of functional behaviour that has been experienced for each pixel, this is described 
in detail in Appendix B.}
\label{WLnumeric}
\end{figure}
Fig. \ref{WLnumeric} uses the diagnostic measure described in Appendix B 
to show that on the scale of $\Delta z=0.2$ every 
possible functional behaviour has been experienced at every point within the bounded region, 
for this example we use $m_0=1\times 10^{-3}$ 
and $\beta=1.0$.

To begin we will first consider $m(z)$ with $m_0=1\times 10^{-3}$ and $\beta=1.0$. Using the 
functional form filling technique we have identified which systematic functions cause 
the largest 
bias for each cosmological parameter. Table \ref{maxbiastab} shows the maximum bias  
to marginal error ratio for each cosmological parameter due to this
particular $m(z)$. 

We have taken into account all degeneracies between parameters in this
exercise -- in the terminology of equation (\ref{biase}) the inverse
Fisher matrix is marginalised over all parameters and the $B_j$ takes
into account degeneracies between the cosmology and systematic functions.

\begin{table}
\begin{center}
\begin{tabular}{|l|c|c|}
\hline
Parameter&bias/marginal error\\
\hline
 $\Omega_m$&$-0.837$\\
 $\Omega_{DE}$ &$ 0.813$\\
 $\Omega_B$&$-0.481$\\
 $h$&$-0.483$\\
 $w_0$&$ 1.253$\\
 $w_a$&$-0.999$\\
 $\sigma_8$&$ 0.855$\\
 $n_s$&$ 0.868$\\
\hline
\end{tabular}
\caption{The ratio of bias to marginal error for 
each cosmological parameter for the fiducial weak lensing survey and a multiplicative systematic 
of the form given in equation (\ref{WL5}) with $m_0=1\times 10^{-3}$ and $\beta=1$. A negative value 
implies that the bias is negative.}
\label{maxbiastab}
\end{center}
\end{table}
\begin{figure}
\resizebox{84mm}{!}{
\includegraphics[clip=true]{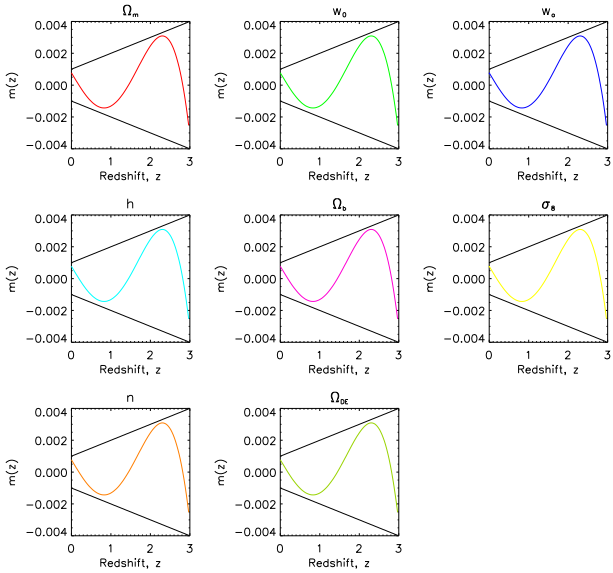}}
\caption{This plot shows the bounded area defined using equation (\ref{WL5}) 
with $m_0=1\times 10^{-3}$ and $\beta=1.0$, shown by the solid black lines. 
Plotted within this area are the functions, 
out of all the possible functions that could be drawn within the bounded region, 
that cause the largest bias on each cosmological parameter, denoted by the panel title.}
\label{worstfuncs}
\end{figure}
Fig. \ref{worstfuncs} shows the worst functions, that cause the largest bias, 
for each cosmological parameter. Since there are degeneracies between all parameters 
the worst function is practically the same for all parameters.

We find that the ratio of the maximum bias to the marginal error is $< 1$ for most 
parameters, except $b(w_0)/\sigma(w_0)=1.25$ which is is not 
acceptable: the bias is larger than the statistical error. 
These results are in rough agreement with Amara \& Refregier (2007b) where it was concluded, 
with a restricted functional parameterisation of $m(z)$, and the assumption of a 
flat Universe, that $m_0=1\times 10^{-3}$ and $\beta=1$ yielded biases that were 
$b(\theta_i)/\sigma(\theta_i)< 1.0$ but were somewhat on 
the limit of what is acceptable. Furthermore it was concluded that functions which 
have a variation (cross from positive to negative) about $z\gs 1$ 
have the largest effect, we also find that 
the functions that cause the largest bias 
in all the parameters varies in the region of $z\sim 1$. This is because it is at $z\sim 1$
that dark energy begins to dominate and so the parameters $w_0$ and $w_a$, 
and through degeneracies the other parameters, are affected by systematic 
functions that introduce variation at this redshift. 

\begin{figure}
\resizebox{84mm}{!}{
\includegraphics{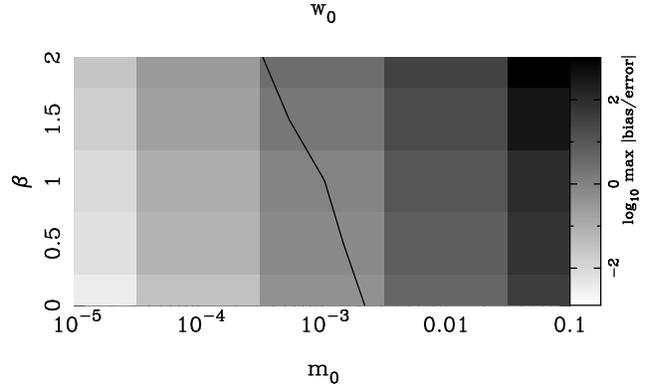}}
\caption{The ratio of maximum bias, found using functional form filling, 
to marginal error as a function of $m_0$ and $\beta$ for $w_0$. The 
gray scale represents the bias to error ratio with a key given on the side of each panel. The solid 
lines show the $b(\theta_i)/\sigma(\theta_i)=1$ contours for each parameter.}
\label{2Dmb}
\end{figure}

Fig. \ref{2Dmb} shows the ratio of bias 
to marginal error as a function of $m_0$ and $\beta$ for $w_0$. 
It can be seen that the redshift
scaling $\beta$ has the expected effect on the maximum systematic bias: as $\beta$ 
increases the maximum bias for a given $m_0$ also 
increases as the systematic bounded area expands. 
The solid lines in Fig. \ref{2Dmb} show the $b(\theta_i)/\sigma(\theta_i)=1$ contours. For 
$m_0=1\times10^{-3}$ we need a $\beta\ls 0.70$ for the bias on $w_0$ to be acceptable. 
If the redshift
scaling is eliminated $\beta=0$ then the requirement on the absolute magnitude of 
$m(z)$ is relaxed to $m_0\ls 2\times 10^{-3}$. 

\begin{figure}
\resizebox{84mm}{!}{
\includegraphics{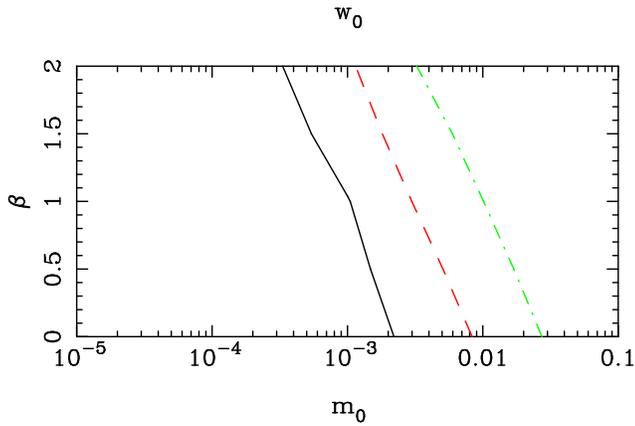}}
\caption{The solid lines show the $|b(w_0)/\sigma(w_0)|=1.0$ contours in the 
($m_0$, $\beta$) parameter space for varying survey area. Black (solid) 
shows the contour for the fiducial survey with Area$=20000$ square degrees, 
red (light gray, dashed) shows the contour for a survey exactly 
the same as the fiducial survey expect 
that Area$=2000$ square degrees, and green (lightest gray, dot-dashed) for a survey with 
Area$=200$ square degrees. For the $w_0$ constraint to be robust to $m(z)$ systematics the values 
of $m_0$ and $\beta$ must lie leftward of the contours, see Fig. \ref{2Dmb}.}
\label{makebiasplots}
\end{figure}
Fig. \ref{makebiasplots} shows the affect of survey area on the requirement of 
$m_0$ and $\beta$. The lines in this figure show the 
$|b(w_0)/\sigma(w_0)|=1.0$ contours for varying survey area. We 
have picked $w_0$ as an example since this parameter 
provides the most stringent constraints on the shape measurement 
parameters (see Table \ref{maxbiastab} and Fig. \ref{2Dmb}). 
As the survey area increases and the marginal error on $w_0$ 
decreases the requirement on shape measurement accuracy becomes more 
stringent. We have fitted a simple scaling relation 
to these contours so that for statistical errors to be reliable the 
following relation holds
\be
0.17\left(\frac{m_0}{1\times 10^{-3}}\right)^{2.4}\left(\frac{\rm Area}{20000}\right)^{1.5}10^{\beta}\leq 1.
\ee

The requirements on $m_0$ and $\beta$ for the largest survey considered 
are at the limit of currently available shape measurement techniques, 
that yield on average  
$m\sim 2\times 10^{-3}$ at best (Miller et al, 2008; Kitching et al., 2008b).
A large redshift variation 
of $m(z)$ results in large biases (Fig. \ref{worstfuncs}) so a 
shape measurement method that is unaffected by the magnitude/size of the 
galaxy population should yield robust cosmological constraints. For 
example Kitching et al. (2008b) have shown that the {\tt lensfit} method has a 
characteristic $m$ that 
has a small magnitude/size dependence. In this case, $\beta\ll 1.0$, 
and a more relaxed constraint 
on $m_0\sim 2\times10^{-3}$ -- $4\times10^{-3}$ is required 
which is well within reach of 
these most recent developments in shape measurement. 
These results are also in agreement with Semboloni et al. (2008) 
where they find that 
for a very restrictive class of $m(z)$ functions, of the form $m(z)=az+b$, 
but a more complex likelihood description, that some 
parameter combinations of $a$ and $b$ can give rise to biases that are 
less that the cosmological errors.

For the smaller surveys considered, that are well matched to currently 
available or upcoming experiments (e.g. CFHTLS, van Waerbeke et al., 2001; 
Pan-Starrs, Kaiser, 2002), 
the shape measurement techniques currently available have biases 
$m$ (Heymans et al., 2006; 
Massey et al., 2007; Kitching et al., 2008b) that are well within 
the required level of accuracy. The caveat to these conclusions is that here 
we have not discussed the size or galaxy-type dependence of any bias and we have not 
considered any calibration offset in the measured shear. 


\section{Conclusion}
\label{Conclusion}

In this article we have presented a method that allows one to move beyond the 
tendency to treat a systematic effect as a statistical one. 
When a systematic is treated as a statistical 
signal extra parameters are introduced to describe the effect, and then these extra nuisance 
parameters are marginalised over jointly with cosmological parameters. We have 
shown that even in 
a very simple toy model such an approach is risky at best. 
The results being highly dependent 
on the choice of parameterisation, and in the limit of a large number of parameters any 
statistical signal on cosmological parameters can be completely diluted. 
One could use marginalisation if there exists a 
compelling physical theory for the systematic, however  
if the functional form is merely phenomenological or if it contains a 
truncated expansion then one should use caution. Even in the case that confidence is high with 
respect to the assumed functional form any residual must be investigated correctly.

As an alternative we advocate treating a systematic signal as such, 
an unknown contaminant in the 
data. 
We address the situation where we have at least some extra information 
on the systematic, 
either from some theory or simulation that may provide a hard boundary within which a 
systematic must lie, or 
from some external data set. We leave the case of `entangled systematics' (where there is no 
extra information and where the systematic depends on the cosmological parameters) 
to be investigated in Amara et al. (in prep). Throughout we have introduced 
each concept using a simple toy model. 

Since the systematic is treated as a genuine unknown we must address every 
possible functional form 
that is allowed, by either the hard boundary or the extra systematic data. 
To do this we use complete basis sets and 
randomly sample from the space of coefficients until all 
functional behaviour has been experienced 
at every point with the hard boundary, or within a few $\sigma$ of the data. 
We have shown that 
such ``functional form filling'' can be achieved by this technique using either Chebyshev, 
Fourier or tophat basis sets. 
By treating each function as a possible systematic, a 
bias in the maximum likelihood value is introduced whilst the marginal error stays the same.
Throughout we have shown that as long as 
functional filling is achieved all conclusions on the magnitude of a systematic effect are 
independent of the choice of complete basis set -- though binning is highly 
inefficient in terms of computational time, and could be labelled as an unphysical basis set.

For the case of a hard boundary every function is given an equal probability and as 
such a maximum 
bias exists. For the case of extra systematic data we show that a 
probability can be assigned to 
each function, and bias, allowing for the production of robust 
systematic probability contours.

We have made a first application of hard boundary functional form filling by addressing 
multiplicative systematics in cosmic shear tomography; 
we have left an application of the data  
bound for future work. We address a lensing systematic, that can 
result from shape measurement or PSF reconstruction inaccuracies, that has the 
same overall shape as the lensing power spectrum but is multiplied by some extra function. 
This is commonly represented  
using a multiplicative function $m(z)$ (Heymans et al, 2006; Massey et al., 2007)
where $\gamma^{\rm sys}=m(z)\gamma^{\rm true}$. 
For a DUNE/Euclid type survey, we find that in order for the systematic 
on $w_0$ to yield a bias smaller than 
the marginal error the overall magnitude of $m(z)$ should be $m_0\leq 8\times 10^{-4}$ 
for a linear
scaling in $(1+z)$, but as the magnitude of the redshift scaling is relaxed then the 
requirement on $m_0$ increases to $m_0\leq 2\times 10^{-3}$. 
The most recent shape measurement methods have $m\sim 2\times 10^{-3}$ (e.g. 
using the {\tt lensfit} method Miller et al., 2007; Kitching et al., 2008b) and have a small 
scaling as a function of magnitude/size. The results shown here then, coupled with some 
currently available shape measurement techniques imply that constraints on cosmological 
parameters from tomographic weak lensing surveys 
should be robust to shape measurement systematics (with the caveat that PSF calibration 
and galaxy size dependence of the bias have been neglected here).

The techniques introduced here should have a wide application in cosmological 
parameter estimation: 
in any place in which there is some signal with some extra information on a systematic. For 
example baryon acoustic oscillations and galaxy bias, CMB and foregrounds, galaxy clusters 
and mass selection, supernovae and light rise times, 
weak lensing and photometric redshift uncertainties. 
A sophistication of these techniques could assign different 
weights/prior probabilities to particular functional forms, 
that are known \emph{a priori} to have a large/small 
effect on cosmological parameter determination, 
such weights could come from either theoretical or instrumental constraints. 

Cosmology is entering into a phase in which the statistical accuracy on parameters will 
be orders of magnitude smaller than anything achieved thus far. 
But we must take care of systematics in a rigorous way 
to be sure that our statistical constraints are valid. 
There exists the pervading worry that the 
functional forms used to parameterise systematics are not representative of the true 
underlying nature of the systematic and that something may have been missed. 
In such a scenario one should always add 
the warning that ``cosmological constraints 
are subject to the assumption of the systematic form''. 

Here 
we have presented a way to address systematics in a way that requires no external assumptions, 
allowing for robust and rigorous statements on systematics to be made. 


\section*{Acknowledgments}
TDK is supported by the Science and Technology Facilities Council,
research grant number E001114. AA is supported by the Swiss 
Institute of Technology through a Zwicky Prize
Fellowship. FBA is supported by an Early Careers Fellowship from 
The Leverhulme Foundation. BJ is supported by the Deutsche Telekom
Stiftung and by the Bonn-Cologne Graduate School of Physics and Astronomy. 
We thank Andy Taylor for many detailed discussions on functional form
filling.   
We thank the organisers and participants 
of the Intrinsic Alignments 
and Cosmic Shear Workshop at UCL on $31^{\rm st}$ March -- $4^{\rm th}$ April 2008. In 
particular Sarah 
Bridle, Catherine Heymans, Rachel Mandelbaum, Lindsay King, Bjoern Schaefer and Oliver Hahn. 
We would also like to thank all members of the DUNE/Euclid weak lensing 
working group. We also thank Eric Linder, Lance Miller, 
Anais Rassat and a rigorous referee 
for many insightful discussions.


\section*{Appendix A : Marginalisation vs. Bias}

Here we will show how marginalisation is mathematically different to the bias formalism used 
in the main article. 

When marginalising the log-likelihood of some cosmological parameters $\btheta$ 
can be written, for a signal $C$ some theory $T$ and a systematic $S$ as 
\be 
\label{MB1}
2{\mathcal L}(\btheta)=\sum_x \sigma_C^{-2}(C-T-S)^2. 
\ee
When marginalising over the systematic the effect is parameterised by some extra parameters 
$\bolda$. If extra data $D$ is provided on the systematic 
then this adds a prior such that the total log-likelihood can be written
\be 
\label{MB2}
2{\mathcal L}(\btheta,\bolda)=\sum_x \sigma_C^{-2}(C-T-S)^2 + \sum_x \sigma_D^{-2}(D-S)^2.
\ee
The Fisher matrix for this is created by taking the derivatives of the log-likelihood with 
respect to  
cosmological parameters $\btheta$ and systematic parameters $\bolda$ so that the first 
and second terms in equation (\ref{MB2}) give the following Fisher matrices 
\ba 
\label{MB3}
F^{\Phib\Phib}&=&
\left( \begin{array}{cc}
 F^{\btheta\btheta} & F^{\btheta\bolda}  \\
F^{\bolda\btheta}   & F^{\bolda\bolda} \\
  \end{array}\right)_C +
\left( \begin{array}{cc}
 0 & 0  \\
 0   & F^{\bolda\bolda}
\end{array}\right)_D\nn
&=& \left( \begin{array}{cc}
 F^{\btheta\btheta} & F^{\btheta\bolda}  \\
F^{\bolda\btheta}   & F^{\bolda\bolda}_C+F^{\bolda\bolda}_D \\
  \end{array}\right).
\ea
The marginal error on the cosmological parameters is found by inverting equation (\ref{MB3}). 
The inverse of the upper left-hand segment of the Fisher matrix in equation (\ref{MB3}) 
can be found by using the Schur complement of the block matrix and then expanding this using 
the Woodbury matrix identity which gives 
\ba 
\label{MSEm}
(F^{\Phib\Phib}_{\rm upper-left})^{-1}&=&{\rm MSE}_{\rm marg}\nn
&=&A^{-1}+A^{-1}B(E-B^{T}A^{-1}B)^{-1}B^{T}A^{-1}\nn
\ea
where $A=F^{\btheta\btheta}$, $B=F^{\btheta\bolda}$ and 
$E=F^{\bolda\bolda}_C+F^{\bolda\bolda}_D$. The marginal error in the case of no systematic ($A^{-1}$) has been increased by an extra factor that depends on the degeneracies between the 
systematic parameters and the cosmological ones ($B$) and on the information available on the 
systematic parameters themselves ($E$). This is also equal to the mean square error of the 
cosmological parameters in this case, since no bias is introduced.

For the bias functional form filling technique the log-likelihood of the cosmological parameters
can again be written as 
\be 
\label{MB4}
2{\mathcal L}(\btheta)=\sum_x \sigma_C^{-2}(C-T-S)^2. 
\ee
In the data bound case we can assign a weight to each systematic \emph{function} $S$ 
(not extra parameters)
\be 
\label{MB5}
2{\mathcal L}(S)=\sum_x \sigma_D^{-2}(D-S)^2 
\ee
so that a very similar expression to equation (\ref{MB2}) can be written for the joint log-likelihood of the cosmological parameters and the systematic
\be 
\label{MB6}
2{\mathcal L}(\btheta,S)=\sum_x \sigma_C^{-2}(C-T-S)^2 + \sum_x \sigma_D^{-2}(D-S)^2.
\ee
We have not necessarily parameterised $S$ with any parameters we wish to measure, $S$ can simply be a function that has been arbitrarily drawn through the systematic data. However if the systematic is parameterised (as we do in the functional form filling approach using complete basis sets) then the signal data $C$ is not used to determine the values of the extra parameters -- 
the number of degrees of freedom in the fit has been reduced with respect to 
the marginalising case. The systematic data $D$ (or a hard boundary) is used to determine the probability of the systematic. Hence the Fisher 
matrix can be written, in this case, like 
\ba 
F^{\Phib\Phib}&=&
\left( \begin{array}{cc}
 F^{\btheta\btheta} & 0  \\
0   & 0 \\
  \end{array}\right)_C +
\left( \begin{array}{cc}
 0 & 0  \\
 0   & F^{\bolda\bolda}
\end{array}\right)_D\nn
&=& \left( \begin{array}{cc}
 F^{\btheta\btheta} & 0  \\
 0   & F^{\bolda\bolda}_D \\
  \end{array}\right).
\ea
The inverse of the upper left-hand segment is now simply $(F^{\btheta\btheta})^{-1}=A^{-1}$. Using the result from Appendix D (for the hard boundary case) a bias is introduced with a maximum value of 
\be 
\label{43}
{\rm max}|b_i|=({\rm const})(A^{-1})_{ij} 
\left[\sum \sigma_C^{-2}\frac{\partial C}{\partial\theta_j}\right]^2=({\rm const})(A^{-1})_{ij} F_{j}^2.
\ee
We have again condensed the notation so that 
$F_j=\sum \sigma_C^{-2}\frac{\partial C}{\partial\theta_j}$. Note that $F\not=B\not=E$. 
Hence the total mean-square error 
on cosmological parameters for the bias case comes from the unaffected marginal error of the cosmological parameter and the bias
\be
\label{MSEb}
{\rm MSE}_{\rm bias}=A^{-1}+{\rm bias}^2=A^{-1}+({\rm const})^2(A^{-1})^2 F^4
\ee
a similar expression can be written for the data bound case, see
Appendix D equation (\ref{Ban}). We have compressed the subscript notation from
equation (\ref{43}) in equation (\ref{MSEb}).

These equations show how marginalisation and the bias formalism relate
in a general sense, and that they are mathematically different
objects i.e.  
when finding the maximum bias using the functional
form filling approach the MSE between marginalisation and functional
form filling \emph{can} be different. 
This difference arises because the space of functions probed by
marginalisation is restricted. In the next section we will show that
if the functional space assessed is the same then the MSE should be equal.

\subsection*{Information Content}

If the parameterisation, and number of (truncated) 
parameters, are the same then the MSE resulting from the bias or
marginalising is the same. We show this here using a simple
illustrative example. 

We approximate the nuisance correlation $C^{\rm sys}$ as a first-order
expansion of a set of parameters 
\be 
C^{\rm sys}(\bolda)=C^{\rm sys}_0+\bolda\dot\nabla_{a}C^{\rm sys}_0.
\ee
This can be though of as a restricted set of functions or a special
case of $C^{\rm sys}$. The nuisance parameters $\bolda$ will be correlated with
the cosmological parameters $\btheta$ and we can form the extended
vector and Fisher matrix as shown in equation (\ref{MB3}), as in
equation (\ref{MSEm}) we can write the marginalised covariance matrix
of $\btheta$ like 
\ba
\lgl \btheta \btheta^t \rgl_c &=& [F_{\Phi\Phi}]^{-1}_{\theta\theta}\nn
&=&F^{-1}_{\theta\theta} + F^{-1}_{\theta\theta}F_{\theta a}
(F_{aa}-F_{\theta a}^t F^{-1}_{\theta\theta}F_{\theta a})^{-1}
F_{\theta a}F^{-1}_{\theta\theta}.
\ea
and there is no bias of the measured parameters. The subscript `c'
on the covariance indicates that we are including the covariance
between $\btheta$ and $\bolda$.

If instead we do not account for the covariance between $\btheta$
and $\bolda$ in the data, the measured error in the parameters is
\be
\lgl \btheta \btheta^t \rgl = [F_{\theta\theta}]^{-1}
\ee
but we have induced a bias in the measurement of $\thetab$, \be
\Delta \btheta = - F_{\theta\theta}^{-1}
\sum \sigma_D^{-2} \frac{\de C^{\rm signal}}{\de \theta}
C^{\rm sys}(\bolda).
\ee
where $\sigma_D^2$ is the variance of the data. Using our linear
expansion of $C^{\rm sys}(\bolda)$ we find
\be
\Delta \btheta = - F_{\theta\theta}^{-1} F_{\theta a} \bolda.
\ee
From this can see the bias effect explicitly comes through the
correlation of the nuisance parameters with the cosmological
parameters, via $F_{\theta a}$.

We can simplify things further by considering a single parameter
$\theta$ and a single nuisance parameter $a$. The
conditional error on the parameter can be simplified to
\be
\lgl \theta \theta \rgl_c = \lgl \theta \theta \rgl (1-r^2)^{-1}
\ee
where we have introduced the correlation coefficient defined as,
\be
r^2 =
F_{\theta\theta}^{-1}F_{\theta a}^2
F_{aa}^{-1}.
\label{corrcoeff}
\ee
We can see this by looking at the inverse of the $2\times 2$
matrix $F_{\Phi\Phi}$;
\be
F_{\Phi \Phi}^{-1} = \frac{1}{F_{\theta\theta}F_{aa}-F_{\theta a}^2}
\left( \begin{array}{cc}
  F_{aa} & -F_{\theta a} \\
  -F_{a\theta} & F_{\theta\theta}F_{aa}
\end{array} \right).
\ee
The marginalized error on $\theta$ is
\be
\lgl \theta \theta \rgl_c
=\frac{F_{aa}}{F_{\theta\theta}F_{aa}-F_{\theta a}^2}
= \lgl \theta \theta \rgl (1-r^2)^{-1}.
\ee
Similarly the marginalized error on $a$ is given by
\be
\lgl aa \rgl_c = \lgl aa \rgl
(1-r^2)^{-1}.
\ee
Note that the marginalized errors on $\theta$
and $a$ go to infinity when the correlation coefficient is
unity, $r=1$. This is because $\theta$ and $a$ are completely
degenerate.

From the definition of the correlation coefficient we can write the
covariance between $\theta$ and $a$ as
\be
\lgl \theta a \rgl_c^2 = \lgl \theta \theta \rgl_c\lgl aa \rgl_c  r^2,
\ee
which agrees with the definition of $r$ given in equation
(\ref{corrcoeff}).

Note that the results for $\lgl \theta \theta \rgl_c$, $\lgl aa
\rgl_c$ 
and $\lgl \theta a \rgl_c^2$ are not just the errors
and covariance when we marginalize. They are also the errors and
covariances when the parameter $\theta$ is correlated with
$a$. The only way to get back the uncorrelated errors is if
$\theta$ and $a$ are uncorrelated, so $r=0$, or if we know
$a$ from some other measurement, in which case they effectively decorrelate. Even if we
assume some value, or range of values, for $a$, there is still a correlation
between possible values of $a$ and $\theta$.

Now consider the bias on $\theta$, $\Delta \theta$, when we ignore
the nuisance effect (or assume some fixed form) . In our simple
model the bias can be written
\be
\Delta \theta = \frac{\lgl \theta \theta \rgl^{1/2}}{\lgl aa
  \rgl^{1/2}} \,a  r =
\frac{\lgl \theta \theta \rgl_c^{1/2}}{\lgl a a
  \rgl_c^{1/2}} \,  a r.
\ee
Now the error on the bias, including all the correlations between
parameters, is given by
\be
\lgl \Delta \theta \Delta \theta \rgl_c
=\left(\frac{\lgl \theta \theta \rgl_c}{\lgl a a
  \rgl_c}\right)    \lgl a a \rgl_c r^2=  \lgl \theta \theta
\rgl_c r^2 = \lgl \theta \theta \rgl \frac{r^2}{1-r^2}.
\label{biaserr}
\ee
One might have mistakenly thought that the covariance between
$\theta$ and $a$ was
\be
\lgl \Delta \theta \Delta \theta \rgl_c
=\left(\frac{\lgl \theta \theta \rgl}{\lgl a a
  \rgl}\right)    \lgl a a \rgl r^2=  \lgl \theta \theta
\rgl r^2,
\ee
but in doing so we have ignored the real correlation between
$\theta$ and $a$ that exists. In the case that of fully
correlated parameters, when $r=1$, the error on $\Delta \theta$ is
only the uncorrelated error on $\theta$, whereas the real
uncertainty is infinite.

Finally the correlation between the bias and the cosmological
parameter is
\be
\lgl \theta \Delta \theta \rgl_c =\frac{\lgl \theta \theta \rgl_c^{1/2}}{\lgl a a
  \rgl_c^{1/2}} \,   r \lgl \theta a \rgl_c
=\frac{\lgl \theta \theta \rgl_c^{1/2}}{\lgl a a
  \rgl_c^{1/2}} \,   r \lgl \theta \theta \rgl_c^{1/2} \lgl a a
\rgl_c^{1/2}  r = \lgl \theta \theta \rgl_c r^2.
\ee

Now we want to compare the effect of the bias with
marginalization on the error of $\theta$. In the case of
marginalization this is given by the correlated covariance
\be
\lgl \theta \theta \rgl_c = \lgl \theta \theta \rgl
(1-r^2)^{-1}.
\ee
In the case of bias, we have to add the uncorrelated
error, $\lgl \theta \theta \rgl$, with the error in the bias
value, $\lgl \Delta \theta \Delta \theta \rgl_c$, which does
include the correlation between cosmological and nuisance
parameters,  to form the MSE;
\ba
MSE &=&\lgl \theta \theta \rgl + \lgl \Delta \theta \Delta \theta
\rgl_c\nn
&=&\lgl \theta \theta \rgl +  \lgl \theta \theta
\rgl_c r^2 =\lgl \theta \theta \rgl +  \lgl \theta \theta
\rgl \frac{r^2}{(1-r^2)}\nn
&=&\lgl \theta \theta \rgl (1-r^2)^{-1}\nn
&=&\lgl \theta \theta
\rgl_c.
\label{MSEant}
\ea
Hence the MSE is the same as the marginalized error in this simple
case, 
and there is
no loss or gain of information. Our result here is for a
2-parameter case, but is easily extended to multiple parameters.

It is worth thinking about the assumptions that lead to this
result. For instance we have assumed in our analysis that the
nuisance bias is \emph{completely specified} by the parameter $\alpha$.

This shows that if one fully understands the nuisance effect,
there is no difference between marginalisation and form-filling.
Given marginalisation can be done quickly, and in some cases
analytically, we would advocate marginalisation in this case.

However, if the form of the nuisance function is not known, or is
wrong, marginalization will tend to underestimate the true systematic
error because of the limited functional space explored by the set of
functions assessed by the truncated basis set.

Given that in general the number of parameters that can be
marginalised over is limited by the number of data points if the
systematic parameterisation is unknown then the marginalising will
necessarily not be able to assess the impact of some functions. 

\subsection*{Constraining nuisance effects with external data}

The effect of additional, external data to constrain the nuisance
effects will add an extra Fisher matrix to  $F_{aa}$;
\be
F_{aa} \rightarrow F'_{aa}=
F_{aa}+F^{\rm ex}_{aa}.
\ee
If we define a new parameter
\be
\beta = [F_{aa}]^{-1}F^{\rm ex}_{aa}
= \frac{\lgl aa \rgl}{\lgl aa\rgl^{ex}}
\ee
as the ratio of the conditional error on $a$ from the
original dataset to the expected measured accuracy on $a$
from the external data. If the error on $a$ from the external
data is small, we can expect $\beta$ to be large, while of the
external error is large, $\beta$ will be small. This results in
the correlation coefficient becoming
\be
r'^2 = \frac{r^2}{1+\beta}.
\ee
When the error on the nuisance parameters is highly constrained by
the new data set, $\beta \gg 1$ and the correlation coefficient
decreases. The effect of new data in the nuisance functions is to
de-correlate the nuisance parameter from the data. Propagating
this through we find
\be
\lgl aa \rgl'_c = \lgl aa \rgl \left(
\frac{1}{1-r^2 + \beta}
\right)
\ee
so that the improved accuracy from the external dataset feeds
through. The marginalised error on the cosmological parameter is
now
\be
\lgl \theta \theta \rgl'_c= \lgl \theta \theta \rgl
\left(\frac{1+\beta}{1-r^2 + \beta}\right).
\ee
The error on the bias is now
\be
\lgl \Delta \theta \Delta \theta \rgl = \lgl \theta \theta
\rgl'_c r'^2 =
\frac{\lgl \theta \theta \rgl r^2}{(1-r^2 + \beta)}.
\ee
Finally, the MSE is
\ba
MSE' &=& \lgl \theta \theta \rgl' + \lgl \Delta \theta \Delta \theta
\rgl'_c\nn 
&=& \lgl \theta \theta \rgl + \frac{\lgl \theta \theta \rgl r^2}{(1-r^2 + \beta)}\nn
&=&\lgl \theta \theta \rgl
\left(\frac{1+\beta}{1-r^2 + \beta}\right).
\ea
Hence adding an external dataset to constrain $a$ does not
change the conclusions that, if the systematic effect can be
modelled by a known parameterization, the MSE is the same as the
marginalized error on $\theta$.

\subsection*{Summary}

We summarise this Appendix by stating the its main conclusions 
\begin{itemize}
\item 
\emph{If} the functional form of the systematic is known then the
degradation in the MSE as a result of marginalising or functional form
filling (bias) is the same.
\item 
If the functional form of the systematic is unknown then the MSE from
marginalising will tend to underestimate the true systematic error and
is in general not equal to the MSE from functional form filling.
\item 
Given that the marginalising necessarily truncates the basis set there
are always some functions that marginalising cannot assess. 
\end{itemize}

\section*{Appendix B: Testing Functional Filling}
In this Appendix we will 
present a numerical 
analysis that will show that if the order and the number of function evaluations is sufficient 
then a bounded area can be exhaustively filled with all possible functional 
forms complete to some 
scale. We want a non-parametric minimum-assumption approach for determining whether we have 
fully sampled the allowed function-space.

\begin{figure}
  \resizebox{84mm}{!}{
    \rotatebox{0}{
      \includegraphics{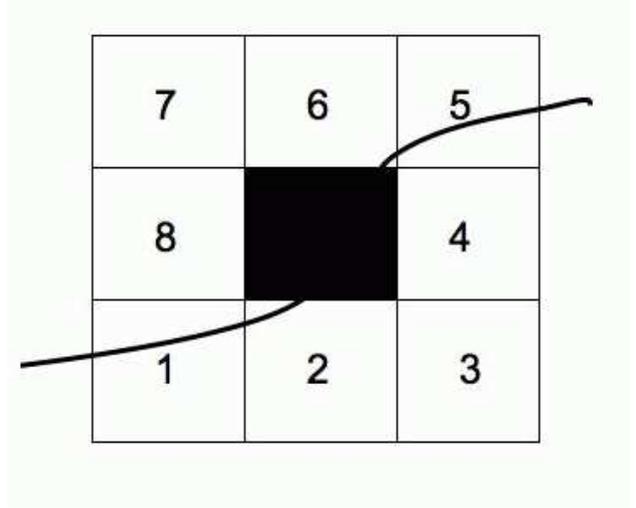}}}
  \caption{For each pixel defined within a bounded region we label the neighbouring pixels. The 
    function shown would assign the functional dependency (enter $2$, exit $6$) to the 
    central pixel.}
\label{A1}
\end{figure}

We can pick a certain scale upon which we can investigate whether all possible 
functional forms are evaluated. Having chosen a scale a bounded area can now be pixelated at 
that scale. In Fig. \ref{A1} we consider a particular pixel and its neighbouring pixels. 
There are $22$ non-degenerate ways in which a function can pass through the pixel in question 
and two surrounding pixels 
(entering via one pixel and exiting via another) 
for example (enter $1$, exit $2$),(enter $1$, exit $3$), 
(enter $1$, exit $4$), ..., (enter $8$, exit $6$). We 
call each combination of entrance and exit
a `functional dependency' or a `functional behaviour'. 
We exclude functions that are multiple-valued
e.g (enter $8$, exit $7$) which would require multiple values of
$y(x)$ for a given $x$. 

Some combinations are more likely than others,
for example consider a box centered on ($0$,$0$). The function $y=x$
will enter through $1$ and exit via $5$, corner functions of this type are
unlikely but not impossible; a more robust and fair measure may
concatenate boxes $1$ and $2$ (and $3$ and $4$) for example. 

We now look at \emph{each} and \emph{every} pixel within the bounded region and for 
each function drawn perform a check of which entrant and exit functional 
dependencies are explored -- checking off each functional dependency as it is sampled. We then 
assign for each pixel the percentage of functional dependencies that have been experienced as 
a result of having drawn the full set of functions. 
We perform this test for the hard boundary toy model presented in Section 
\ref{A Simple Toy Model} for each of the basis sets considered in Section 
\ref{Functional Form Filling}, Chebyshev, Fourier and tophat functions. 
We increase the maximum 
order $N$ in the expansion in equation (\ref{E9}) and the number of random realisations $N_F$ 
of the coefficient space $\{a_0$,...,$a_N\}$ (and $\{b_0$,...,$b_N\}$ for the Fourier basis set) . 

\begin{figure}
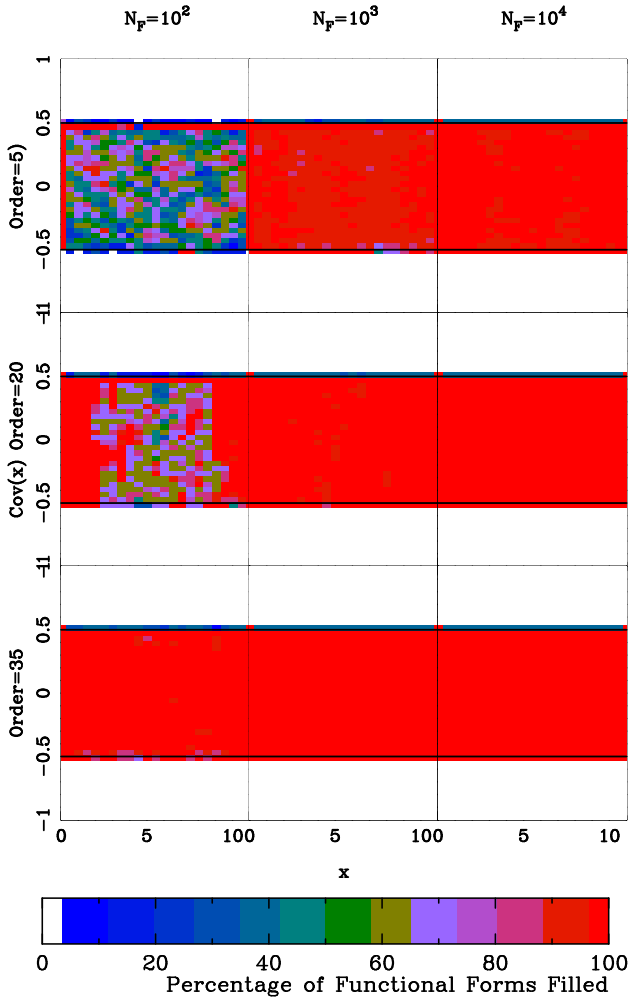

  \resizebox{84mm}{!}{
    \rotatebox{0}{
      \includegraphics{numericCheb.ps}}}
  \resizebox{84mm}{!}{
    \rotatebox{0}{
      \includegraphics[clip=true]{key.eps}}}
  \caption{For the {\bf Chebyshev} basis set. For 
    each pixel defined within the simple bounded region we sum up all functional 
    dependencies experienced from the full set of functions evaluated.}
\label{A2}
\end{figure}

\begin{figure}
  \resizebox{84mm}{!}{
    \rotatebox{0}{
      \includegraphics{numericFour.ps}}}
  \resizebox{84mm}{!}{
    \rotatebox{0}{
      \includegraphics[clip=true]{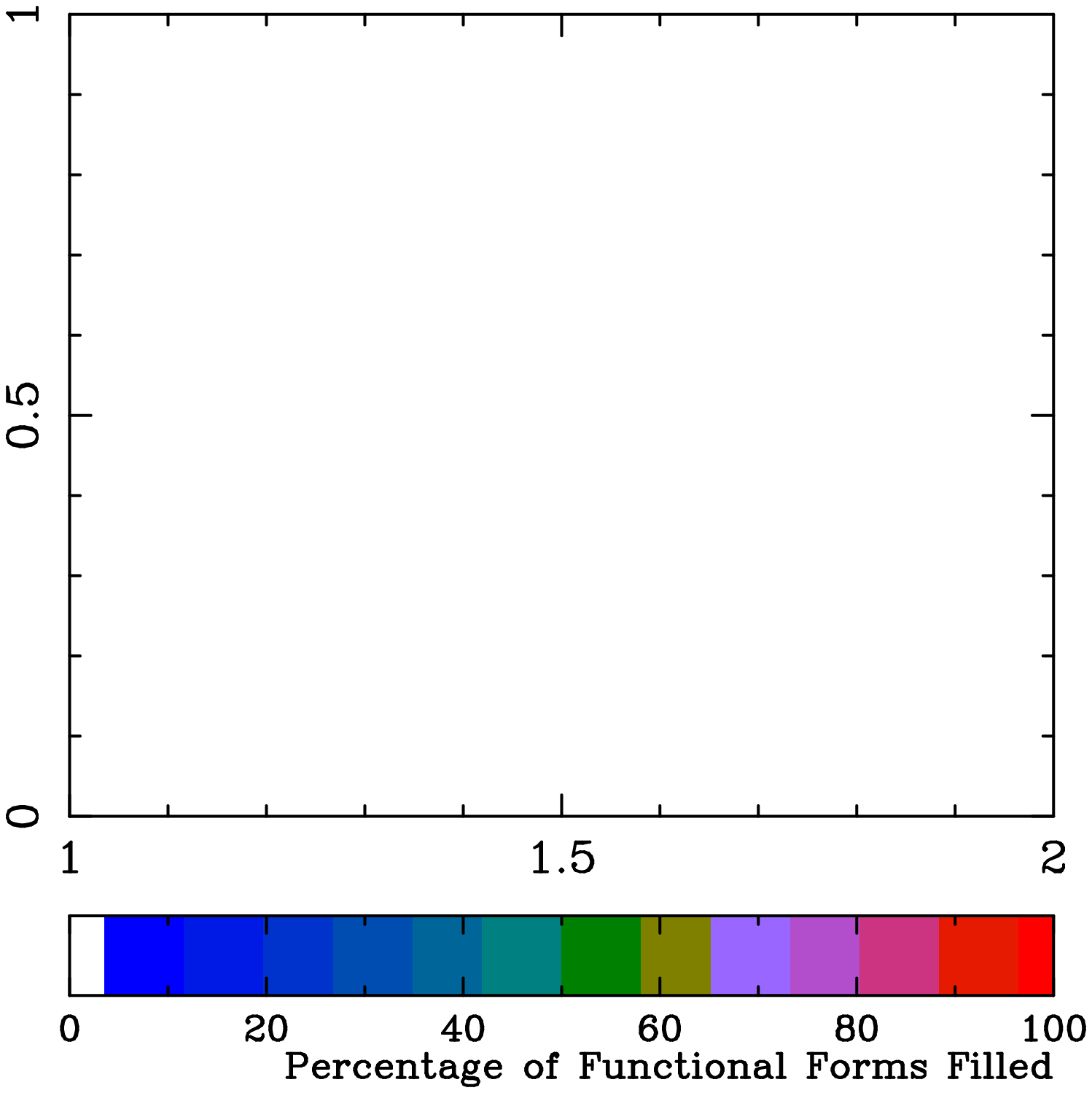}}}
  \caption{For the {\bf Fourier} basis set. For 
    each pixel defined within the simple bounded region we sum up all functional 
    dependencies experienced from the full set of functions evaluated.}
\label{A3}
\end{figure}

\begin{figure}
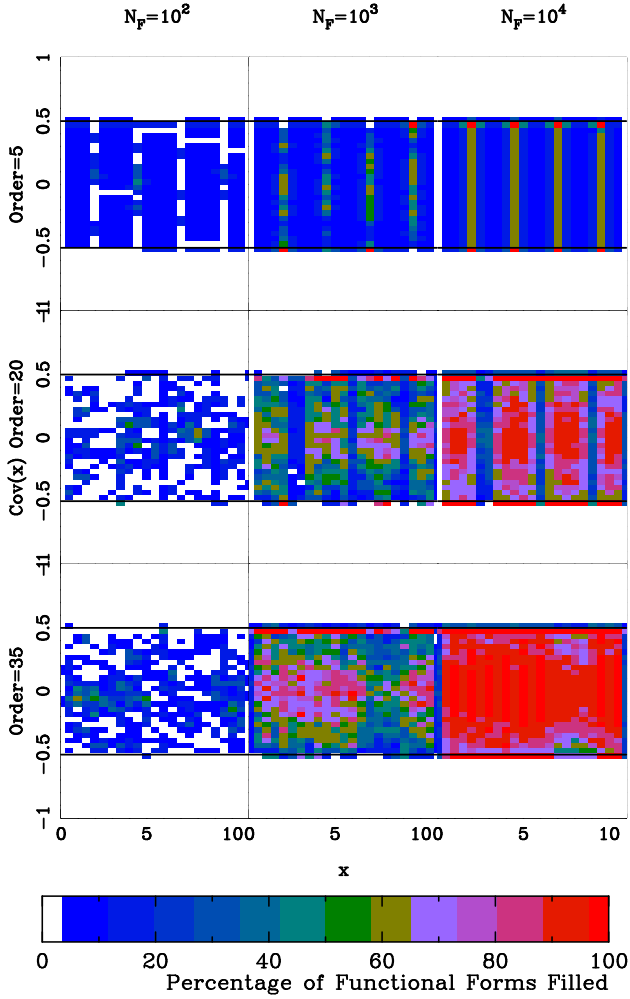

  \resizebox{84mm}{!}{
    \rotatebox{0}{
      \includegraphics{numericDelta.ps}}}
  \resizebox{84mm}{!}{
    \rotatebox{0}{
      \includegraphics[clip=true]{key.eps}}}
  \caption{For the {\bf tophat} basis set. For 
    each pixel defined within the simple bounded region we sum up all functional 
    dependencies experienced from the full set of functions evaluated. For the tophat 
    basis order corresponds to `number of bins'.}
\label{A4}
\end{figure}

Figs \ref{A2} to \ref{A4} show the toy model hard boundary that has been pixelated on the 
scale of $\Delta x=0.5$. It can be seen that within the bounded area every pixel 
does experience all 
functional behaviour if the order and number of functions drawn is sufficiently large. 
As expected when the functional order increases, and as the number of function evaluations 
increases, the number of pixels that experience all functional dependencies increases. We find 
that a maximum order of $N\gs 35$ with $N_F \gs 10^4$ realisations 
is sufficient for each point in the bounded area to experience every functional form 
for this simple example. 

For the
tophat basis set (binning) we find that the maximum order (number of bins) needs to 
be much larger than for either Chebyshev or Fourier basis sets. This can be understood since 
if the bin width (order) is larger than the resolution of the pixels then it is 
impossible for any given pixel to experience particular functional dependencies (such as 
[enter $1$, exit $2$]). So even though tophat functions can be used 
for functional form filling there is some computational expense in this choice. 

\begin{figure}
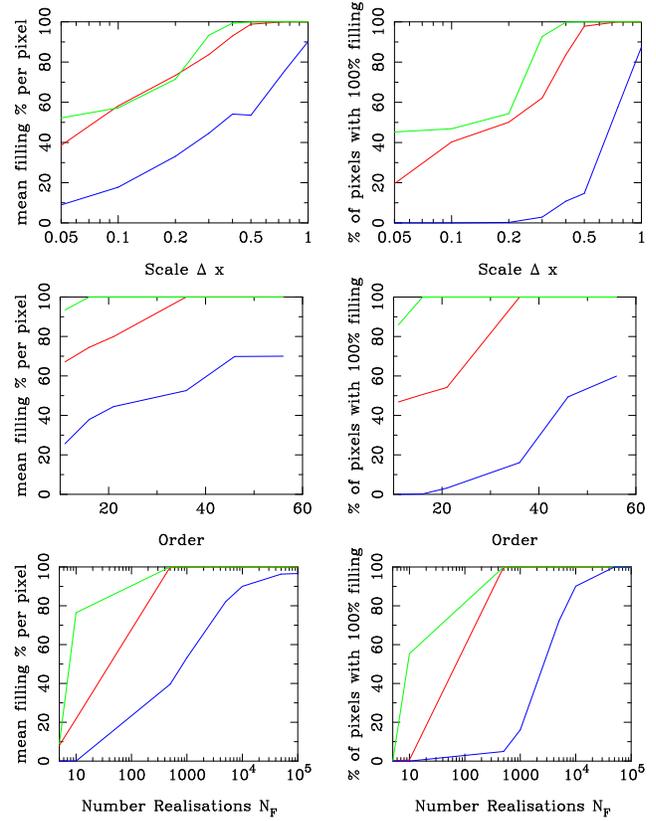

  \resizebox{40mm}{!}{
    \rotatebox{0}{
      \includegraphics{numericScale1.ps}}}
  \resizebox{40mm}{!}{
    \rotatebox{0}{
      \includegraphics{numericScale2.ps}}}
  \resizebox{40mm}{!}{
    \rotatebox{0}{
      \includegraphics{numericOrder1.ps}}}
  \resizebox{40mm}{!}{
    \rotatebox{0}{
      \includegraphics{numericOrder2.ps}}}
  \resizebox{40mm}{!}{
    \rotatebox{0}{
      \includegraphics{numericReali1.ps}}}
  \resizebox{40mm}{!}{
    \rotatebox{0}{
      \includegraphics{numericReali2.ps}}}
   \resizebox{40mm}{!}{
    \rotatebox{0}{}}
  \caption{The right panels show the percentage of pixels in the 
    bounded area that have 
    experienced every type of functional behaviour (100\% filling). 
    The left hand panels show the percentage of all functional behaviours experienced 
    per pixel on average. The upper panels shows these diagnostic measures 
    of functional form filling 
    as a function of scale, with the order and number of
    realisations fixed at $N=35$ and $N_F=10^3$ respectively.
    The middle panels 
    as a function of maximum order, with the scale and number of
    realisations fixed at $\Delta x=0.5$ and $N_F=10^3$ respectively.
    And the lower panels 
    as a function of number of random realisations, with the scale and order fixed at 
    $\Delta x=0.5$ and $N=35$ respectively.
    In all panels blue (darkest gray, lower lines) is for the 
    tophat (binning) basis set, red (lighter gray) for the Chebyshev basis set and 
    green (lightest gray) for the Fourier basis set.}
\label{A5}
\end{figure}
Fig. \ref{A5} shows some diagnostic plots representing 
the completeness of the functional space sampled.  
We show the percent of pixels 
that have 100\% of the possible functional behaviours sampled, and the percentage of 
functional behaviours sampled by the average pixel.
In the top panels we vary the scale that is investigated, and  
it can be seen that for the Chebyshev and Fourier basis sets 
$\sim 100\%$ of pixels have experienced every functional behaviour down to a scale of 
$\Delta x\approx 0.5$. 
The tophat basis set performs much worse with a mean filling of $\sim 10\%$, 
and only $\sim 50\%$ of all pixels experiencing every possible behaviour. 
The general trend is that as the scale drops below the average oscillation 
length of the most highly varying functions the percentage of `good' pixels (experiencing all 
functional behaviour) sharply declines. The tophat basis set fails at $\Delta x\sim 0.3$ since,
with $N=35$ this is approximately the `bin width' of the functions. 
In a cosmological application 
one would ensure that the form filling was complete down to the scale upon which a particular 
parameter affects the signal. 

In the middle panels of Fig. \ref{A5} we vary the maximum order of the 
basis set expansion whilst keeping both the scale and the number of random 
realisations fixed. It can 
again be seen that for a maximum order $\gs 35$ the Chebyshev and Fourier 
basis sets achieve 100\% 
functional filling. Again the 
tophat basis set (binning) fails to achieve any complete functional filling, 
and only begins to fill 
in some pixels when the number of bins becomes more than the number of pixels. 

The lower panel of Fig. \ref{A5} shows how the filling efficiency varies with the number of 
random realisations of the basis set. For the simple example given the 
Chebyshev and Fourier basis 
sets completely fill the bounded area, down to the scale of $\Delta x=0.5$ 
in $\sim 10^3$ realisations whereas the tophat basis set 
requires many orders of magnitude more realisations. 

One could imagine futher metrics that could be used
to gauge the completeness of the functional space 
that has been sampled, in this first exposition of the methodology we have 
shown a simple way to gauge this effect. 

One concern is that this metric may favour the tophat basis since we are using 
a pixelated measure of scale. However we find that even with this advantage  
the tophat basis set achieves complete filling only at the 
expense of a prohibitive amount of computational time compared 
to the Chebyshev and 
Fourier basis sets, as the order and the number of realisations would need to be 
much larger to compensate for the pathological choice of basis functions.

We note that a step-function (e.g. function $f$ over the
interval $x\in[0,1]$which is equal to $+1$ for $0<x<0.5$ and $-1$ for $0.5<x<1$) 
may be better approximated by a low order tophat function and may take a very 
high order Chebyshev or Fourier expansion. 
The feature highlighted here -- a step function -- has a feature (the step) which 
has a very small scale variation. In fact the scale of a step is actually zero. 
The issue of wether functional form filling is complete down to some scale 
is shown in Fig. \ref{A5}, and as the maximum order is increased and more highly 
oscillatory functions are included the minimum complete scale will decrease. 
This example highlights that fact that the tophat basis set can 
probe these very small scales to some degree whilst missing larger scale 
features, this can be seen in Fig. \ref{A4}.

For all the calculations in Section \ref{Functional Form Filling} 
we pessimistically use a maximum order of $N=35$ and 
use $N_F=10^4$ realisations for all basis sets.

\section*{Appendix C : Computational Time} 

One may be concerned at the amount of computational time that the 
functional form filling we are 
advocating may take. This is a valid convern since many realisations 
of the systematic data 
could be required and one must re-perform the cosmological 
parameter fitting for each and every possible function in order to 
evaluate the potential bias. 

If, for a simple grid search in parameter space, 
the time taken to analyse a single point in parameter space is $\tau$ then for $N$ 
cosmological parameters the total time scales as $T\propto A^N\tau$ where 
$A$ is the number 
of evaluations per parameter. Even for one parameter $A\gg 1$ to correctly map 
out a likelihood surface. If one marginalises over 
$M$ extra systematic parameters then the total time must increase to 
$T_{\rm marg}\propto A^{(N+M)}\tau$. 
For the functional form filling case the total time taken for the 
calculation is simply the 
number of different function evaluations $F$ multiplied by the number 
of data realisations $D$ 
and the time to estimate the cosmological parameters is $T_{\rm fff}\propto FDA^N\tau$. 
The ratio in the computation time can now be written as 
$T_{\rm marg}/T_{\rm fff}\approx A^M/FD$. 

Let us pessimistically 
assume that approximately $F=10^4$ function evaluations are needed and $D=10^3$ data 
realisations, and conservatively assume 
that $M=10$ extra systematic parameters are required to ensure a 
systematic effect is correctly 
determined. For $A\gs 100$ we find $T_{\rm marg}\gs 10^9T_{\rm fff}$. 
So on the contrary to such a technique being computationally expensive, 
treating systematics 
in such a fashion may in fact be much more efficient than marginalising 
over many nuisance parameters using a traditional grid search.

For some Monte Carlo parameter searches the computational time can be 
reduced to $T=NA\tau$. In this 
case we find that $T_{\rm marg}/T_{\rm fff}\approx M/NFD+1/FD$. 
The computation time is now longer for functional form filling 
$T_{\rm marg}/T_{\rm fff}\ll 1$. This 
highlights the paramount importance of finding efficient form filling 
functions such as Chebyshev 
polynomials; we investigate the efficiency of Chebyshev, Fourier and 
tophat basis sets in Appendix B.

We take random realisations of the coefficient-space to uniformly sample the 
coefficients' possible values. This is a first, brute-force attempt at the problem. 
Alternative approaches could explore a 
set grid of values in this space, or use a more sophisticated 
random search such as a Monte Carlo chain approach -- we leave this for future work. 


In this first investigation we have taken a simplistic approach 
and, as shown in Appendix B, 
we have found that a set of simple uniform random realisations is sufficient 
to explore the full space of functions.

{\bf Bayesian vs. Frequentist} One may be concerned that what we are advocating is a frequenist 
solution to the systematic problem. On contrary what we suggest is explicitly Bayesian : the 
method can take into account any prior information on the systematic from either theory or data. 
The specific method used in this article to find the form filling functions is analogous to a 
brute force parameter in maximum likelihood rescontruction. 

\section*{Appendix D : Extremal of the Bias}

Here we will show that given a hard boundary or some data there exists a maximum 
absolute bias. Within the Fisher matrix formalism the 
bias caused by a function can be written (equation 
\ref{biase}) as 
\be 
b(\theta_i)=(F^{-1})_{ij}\sum s_C^{-2} C^{\rm sys}\frac{\partial C^{\rm signal}}{\partial \theta_j}
\ee
where $s_C$ is the observed signal variance. We can rewrite this as 
\be
b_{\mu}=(F^{-1})_{\mu\nu}C^{\rm sys}_{\alpha}s^{-2}_{\alpha}D_{\alpha,\nu}
\ee
where $D_{\alpha,\nu}\equiv \frac{\partial C^{\rm signal}_{\alpha}}{\partial \theta_{\nu}}$, we
compress this to 
\be
b_{\mu}=C^{\rm sys}_{\alpha}Q_{\alpha;\mu}
\ee
where $Q_{\alpha;\mu}=(F^{-1})_{\mu\nu}s^{-2}_{\alpha}D_{\alpha,\nu}$.

We have the additional constraint that we are considering the set of systematic functions 
$\{S\}$ that give the same weight 
\be
\label{const}
\sum_{\alpha}\sigma_{\alpha}^{-2}(C^{\rm sys}_{\alpha}-d_{\alpha})^2=A^2={\rm constant}
\ee
with respect to some data vector $d_{\alpha}$. In this proof and
throughout the article 
we assume that there is at least some systematic information at 
every data value in te signal. If there was no constraint at the
position of some of the signal data values then the bias would be unbounded.

For the hard boundary there is a  
constraint that $[C^{\rm sys}(x)]^2 \leq A^2$ at all $x$ 
-- this is actually for a constant 
(flat) hard boundary for a variable boundary the $A\rightarrow A(x)$.

So to find the maximum bias we need to solve the following equation 
\be
\label{53}
\frac{\partial}{\partial C^{\rm sys}_{\gamma}}\left\{C^{\rm sys}_{\alpha}Q_{\alpha;\mu}-
\lambda\left(\sum_{\alpha}\sigma_{\alpha}^{-2}(C^{\rm sys}_{\alpha}-d_{\alpha})^2-A^2\right)\right\}=0
\ee
where $\lambda$ is a Langrange multiplier. Solving this for fixed $\mu$ and all $\gamma$ 
we find that 
\be
C^{\rm sys}_{\gamma}=\frac{1}{2\lambda}\sigma_{\gamma}^{2}Q_{\gamma;\mu}+d_{\gamma}.
\ee
To determine the value of the Lagrange multiplier we substitute this back into the constraint 
equation (\ref{const}) to get a quadratic in $(1/2\lambda)$
\ba
\left(\frac{1}{2\lambda}\right)^2\left(\sum_{\alpha}(\sigma_{\alpha}^{2}Q_{\alpha;\mu})^2\right)+
\left(\frac{1}{2\lambda}\right)  \left(\sum_{\alpha}(2 \sigma_{\alpha}^{2}Q_{\alpha;\mu}d_{\alpha})\right)+\nn
\left(\sum_{\alpha}d^2_{\alpha}-A^2\right)=0.
\ea
This has solutions of the form 
\be
\left(\frac{1}{2\lambda}\right)=R_{\mu}\pm T_{\mu}
\ee
where 
\ba
R_{\mu}&=&-\frac{\left(\sum_{\alpha}2\sigma_{\alpha}^{2}Q_{\alpha;\mu}d_{\alpha}\right)}
{2\sum_{\alpha}\left(\sigma_{\alpha}^{2}Q_{\alpha;\mu}\right)^2}\nn
T_{\mu}&=&\frac{\sqrt{
\left[\sum_{\alpha}\sigma_{\alpha}^{2}Q_{\alpha;\mu}d_{\alpha}\right]^2
-4\left[\sum_{\alpha}\left(\sigma_{\alpha}^{2}Q_{\alpha;\mu}\right)^2\left(\sum_{\alpha}d^2_{\alpha}-A^2\right)\right]}}{2\sum_{\alpha}\left(\sigma_{\alpha}^{2}Q_{\alpha;\mu}\right)^2}\nn.
\ea
So the function(s) with the maximum bias can be expressed as 
\be 
C^{\rm sys}_{\gamma}=(R_{\mu}\pm T_{\mu})\sigma_{\gamma}^{2}Q_{\gamma;\mu}+d_{\gamma}.
\ee
Note that there exists two functions that represent the maximum and minimum bias, 
which are given by 
\be
\label{maxmin}
{\rm max/min}(b_{\mu})=\left[R_{\mu}\sigma_{\alpha}^{2}Q_{\alpha;\mu}+d_{\alpha}\right]Q_{\alpha;\mu}\pm\left[T_{\mu}\sigma_{\alpha}^{2}Q_{\alpha;\mu}\right]Q_{\alpha;\mu}. 
\ee

So for a set of systematic functions that give the 
same weight with respect to some data there 
should exists a maximum and a minimum bias. 
Equation (\ref{const}) can be understood by looking at Fig. \ref{ChiExample}.
For the data boundary case there will be some large and small
biases - actually every bias is `allowed' but has some probability with    
repsect to the data. Equation (\ref{const}) 
effectively takes a horizontal cut across on of the scatter plot panels in 
Fig. \ref{ChiExample} at a given weight so that equation (\ref{maxmin}) gives 
the minimum and maximum biases for that weight.

If the mean of the data is zero $\langle d_{\alpha}\rangle=0$ and the variance of the data is 
$\langle d_{\alpha}^2\rangle=\sigma^2_d$ then $\langle R_{\mu}\rangle=0$ and 
\be
\label{Tan}
\langle T_{\mu}\rangle=\frac{\sqrt{A^2-\sigma_d^2}}
{\left[\sum_{\alpha}\left(\sigma_{\alpha}^2Q_{\alpha;\mu}\right)^2\right]^{{\frac{1}{2}}}}
\ee
so that the mean bias is zero and the bias contours (contours are drawn for a 
constant weights, $A$) can be written as  
\be 
\label{Ban}
\langle b_{\mu}\rangle =\pm\langle T_{\mu}\rangle\sigma_{\alpha}^{2}Q^2_{\alpha;\mu}. 
\ee

For a hard boundary we have 
\ba
\label{hhbb}
R_{\mu}&=&0\nn
T_{\mu}&=&\frac{A}{\left[\sum_{\alpha}\left(Q_{\alpha;\mu}\right)^2\right]^{\frac{1}{2}}}
=A'
\ea
which gives solutions for the maximum bias 
\be 
{\rm max/min}(b_{\mu})=\pm A'Q^2_{\alpha;\mu}
\ee
this calculation can be compared to equations (\ref{Tan}) and (\ref{Ban}).
This confirms that in the hard boundary case the absolute value of the bias has a maximum and 
that there exists two mirrored functions ($f(x)$ and $-f(x)$) which both yield  
this maximum absolute bias.

In the case when the
systematic is constrained by some data points $C^sys_\alpha$,
we can easily construct a function that goes through all the systematic data
points but has unbounded ``bad effects'' at some other point.
However such a function, that
has the same fit to the data but some other behaviour between the
points is also assessed in terms of its impact on the cosmological
parameter in question. Looking at Figure (\ref{ChiExample}) such a function would have
the same weight but would have a smaller bias than a ``good function'' that
went through the same data points because the $a_0$ behaviour is not strongly
degenerate with some ``bad behaviour'' but has a simple functional
dependency. 

The key conclusion of this Appendix is that out of the
space of all functions that have the same weight with respect to the
data there exists a maximum and a minimum bias that this space of
functions can cause.

\section*{Appendix E: Data Weighting}

We know that given some error bars on data, and a different realisation of 
the experiment, that the actual data points would be scattered differently 
but the statistical spread of the data would be the same. If a line 
fits \emph{exactly}
through some data points it is given a likelihood of exactly $1$ but we know that this
probability is spurious since given another realisation of the experiment the data 
would be differently scattered and a new function would be assigned a probability 
of 1. This is a problem because $P=1$ should be unique. This problem occurs because 
in the original step the probability of the data has not been taken
into account.

For the specific
Gaussian case only we have used the marginalisation over the probability of
the data can be done analytically. We stress here that in general the
systematic mean will be non-zero and that \emph{only} in this very
simplified Gaussian case (and some other analytic examples) can this
procedure can ve done anlytically. 

Here we will outline how this `sample variance effect' -- 
that of taking into account the 
likelihood of the systematic data -- 
can be incorporated into the $\chi^2$ weighting scheme given in 
equation (\ref{E10}).

We can write the probability of a function $f[\bolda]$ as the sum over 
the data vector $d_i$ 
multiplied by some prior probability of the data values $p(d_i)$ 
\be
\label{B2}
p(f[\bolda])=\prod_i p(f_i|d_i) p(d_i) 
\ee
where $f_i=f(x_i;\bolda)$ are the function values are the variable positions $x_i$ at which 
the data has been taken.

We now want to marginalise over the probability of each data point to obtain $p(f[\bolda])$
\ba
\label{B3}
p(f[\bolda])&=&\int p(f[\bolda]|D) {\rm d}D\nn 
&=& \prod_i \int  p(f_i|d_i) p(d_i) {\rm d}d_i.
\ea
Since we assume Gaussian data throughout the probability of the data can be written as a 
Gaussian centered about zero and $p(d_i|f_i)$ can be replaced with a $\chi^2$ 
distribution such that 
\be
\label{B4}
p(f[\bolda])\propto\prod_i \int_{-\infty}^{+\infty} 
{\rm e}^{-\frac{(d_i-f_i)^2}{2\sigma_i^2}}{\rm e}^{-\frac{d^2_i}{2\sigma_{D}^2}} {\rm d}d_i.
\ee
where $\sigma_i$ is the error on the $i^{\rm th}$ data 
point and and $\sigma_D$ is the variance of 
data which we take to be $\sigma_D=\sigma_i$. Note that each data point 
could have a different error bar - we have made the assumption that the 
data at each point is Gaussian distributed not that all data points are 
the same (the integral is over $d_i$ not over $i$). 
Evaluating the integral in equation (\ref{B4}) we have (now using log-likelihood for clarity)
\be
\label{B5}
\ln(p(f[\bolda]))\propto -\sum_i\left(\frac{f_i^2}{4\sigma_i^2}\right)
+\sum_i\ln(\sigma_i\sqrt{\pi}).
\ee
So that in the case of Gaussian distributed data the probability of each function (and hence 
the probability of the bias incurred by that function) can be simply evaluated using 
equation (\ref{B5}).
Note that a Gaussian with a mean of zero is unqiuely 
defined by its variance hence the data values themselves do not appear in the 
weighting formula given. 

\section*{Appendix F : iCosmo module description}

The `worst bias' calculations presented in Section 
\ref{An Application to Cosmic Shear Systematics} were done using an extension to 
the open source interactive cosmology calculator {\tt iCosmo} 
(Refregier et al., 2008b; {\tt http://www.icosmo.org}). This additional module 
will be included in v$1.2$ and later. 

The tomographic lensing module {\tt mk\_bias\_cheb} takes $m_0$ and $\beta$ 
defined in equation (\ref{WL5}) and uses the functional form filling technique 
(using the Chebyshev basis set) to calculate the maximum bias in each cosmological 
parameter. The lensing survey and central cosmology can be arbitrarily defined 
using the common {\tt set\_fiducial} routine described in Refregier et al. (2008b).

\end{document}